\documentclass{aa}

\usepackage{graphicx}
\usepackage{txfonts}
\usepackage{color}
\usepackage{listings}
\usepackage{array}
\usepackage{subfigure}
\usepackage{hyperref}
\usepackage{comment}

\newcommand{\GravC}{\mathcal G} 
\newcommand{\redmass}{\mathfrak{m}}
\newcommand{\GM}{\mu} 
\newcommand{\mEarth}{M_\oplus}

\newcommand{\disk}{\mathrm{disc}}
\newcommand{\pl}{\mathrm{pl}}
\newcommand{\mig}{\mathrm{m}}
\newcommand{\tot}{\mathrm{tot}}
\newcommand{\Lind}{\mathrm{L}}
\newcommand{\Corot}{\mathrm{C}}
\newcommand{\bfGamma}{\boldsymbol{\Gamma}}
\newcommand{\bfr}{\boldsymbol{r}}
\newcommand{\bfv}{\boldsymbol{v}}
\newcommand{\bfa}{\boldsymbol{a}}
\newcommand{\bfF}{\boldsymbol{F}}
\newcommand{\ANGMOM}{\mathcal L}

\newcommand{\D}[1]{\ensuremath{\operatorname{d}\!{#1}}} 

\newcommand{\gas}{\mathrm{gas}} 
\newcommand{\accr}{\mathrm{accr}} 
\newcommand{\Kepl}{\mathrm{K}} 
\newcommand{\alphaSigma}{\alpha_\Sigma} 
\newcommand{\betaT}{\beta_T} 
\newcommand{\betah}{{\beta_\mathrm{f}}} 
\newcommand{\alphaturb}{\alpha_\mathrm{t}} 
\newcommand{\nuvisc}{\nu_\mathrm{t}} 
\newcommand{\mpl}{m_{\mathrm{pl}}}

\begin{document}

   \title{A recipe for orbital eccentricity damping in the type-I regime for low viscosity 2D-discs}


   \author{G. Pichierri
          \inst{1}
          \and
          B. Bitsch
          \inst{1}
          \and
          E. Lega\inst{2}\fnmsep
          }

   \institute{Max-Planck-Institut für Astronomie, Königstuhl 17, 69117 Heidelberg, Germany \thanks{pichierri@mpia.de}
         \and
             Laboratoire Lagrange, Université Cote d’Azur, Observatoire de la Cote d’Azur, 06304 Nice, France\\
             }

   \date{}

 
  \abstract
   {
   It is well known that partial and deep gap opening depends on the disc's viscosity; however, damping formulas for orbital eccentricities have only been derived at high viscosities, ignoring partial gap opening.
   }
   {In this work, we aim at obtaining a simple formula to model eccentricity damping of the type-I regime in low viscosity discs, where even small planets of a few to a few tens of Earth's masses may start opening partial gaps in the gas surface density around their orbit.}
   {We perform high resolution 2D locally isothermal hydrodynamical simulations of planets with varying masses on fixed orbits in discs with varying aspect ratios and viscosities. We determine the torque and power felt by the planet to ultimately derive migration and eccentricity damping timescales.}
   {   We first find a lower limit to the gap depths below which vortices appear; this happens roughly at the transition between type-I and classical type-II migration regimes. For the simulations that remain stable, we obtain a fit to the observed gap depth in the limit of vanishing eccentricities that is similar to the one currently used in the literature but is accurate down to $\alpha=3.16\times 10^{-5}$. We then record the eccentricity damping efficiency as a function of the observed gap depth and the initial eccentricity: when the planet has opened a deep enough gap such that the surface density is less than $\sim 80\%$ of the unperturbed disc surface density, a clear linear trend is observed independently of the planet's eccentricity; at shallower gaps this linear trend is preserved at low eccentricities, while it deviates to more efficient damping when $e$ is comparable to the disc's scale height. Both trends can be understood on theoretical grounds and are reproduced by a simple fitting formula.}
   {Our combined fits for the gap depth and eccentricity damping efficiency yield a simple recipe to implement type-I eccentricity damping in $N$-body codes in the case of partial gap opening planets that is consistent with high-resolution 2D hydro-simulations. The typical error of the final fit is of the order of a few percent, and at most $\sim 20\%$, which is the error of type-I torque formulas widely used in the literature. This will allow a more self-consistent treatment of planet-disc interactions of the type-I regime for population synthesis models at low viscosities.}


   \maketitle
%

\section{Introduction}
After more than 25 years of observations, we now have detected over 5000 exoplanets, and have access to stunning observations of protoplanetary discs thanks to instruments like the Atacama Large Millimeter/submillimeter Array (ALMA). The exoplanet sample, together with our own Solar System, has revealed the existence of planets in our Galaxy of vastly different sizes, orbits and bulk compositions, going from the smallest and densest terrestrial-type planets, to the Super-Earths and Mini-Neptunes with their moderate atmospheres, up to the giant Jupiter-like planets.
In the meantime, high-angular-resolution images have recently shown the existence of detailed substructures in protoplanetary discs, such as rings, gaps and spirals (e.g.\ DSHARP survey, \citealt{2018ApJ...869L..41A, 2018ApJ...869L..42H}, and MAPS program, \citealt{2021ApJS..257....1O,2021ApJS..257...14S}), which challenge our understanding of the structure and evolution of such objects.
Naturally, these two classes of astronomical objects are closely related to each other, so that a crucial, bipartite question is how the protoplanetary disc environment sculpts the forming planetary system and simultaneously how forming planets shape the structure of the disc they are embedded within.

For one, a commonly proposed explanations for the substructures seen in DSHARP discs, such as rings and gaps in the millimeter emission together with corresponding deviations in Keplerian velocity of the gas, is that these features probe planet-disc interactions, thus revealing the ongoing process of planet formation \citep{2012A&A...545A..81P, 2018ApJ...869L..47Z, 2018ApJ...864L..26B,2018ApJ...860L..12T,2018ApJ...860L..13P}; at least in one case, the well-known PDS-70 system, two forming protoplanets have been detected \citep{2018A&A...617A..44K, 2018ApJ...863L...8W, 2019NatAs...3..749H}. Even more so, many of these putative forming planets are found in a region of the orbital-period vs.\ planetary mass planet that is currently unavailable to planet-detection methods used so far, due to observational biases \citep{2019MNRAS.486..453L, 2019MNRAS.488.3625N, 2018ApJ...864L..26B, 2022A&A...663A.163M}. The study of planet-disc interactions is therefore a useful tool to uncover exoplanetary systems from disc observations while they are forming.
On the other hand, one of the main goals of planetary science is to construct a model to predict, at least in a statistical sense, what planetary system will emerge around a given star from a given disc (e.g.\ having given surface density, temperature and thickness profiles, a given level of turbulent viscosity, etc., e.g \citealt{2008ApJ...673..487I, 2009A&A...501.1139M, 2013A&A...558A.109A,
2017MNRAS.464..428A,
2017MNRAS.470.1750I, 
2018MNRAS.474..886N, 2019A&A...623A..88B, 2019MNRAS.486.5690G, 2021A&A...650A.152I, 2021A&A...656A..69E}). Clearly, such a model must take into account disc-planet interactions, and their outcome can vary widely under different underlying disc environments.

Although they are unknown in our Solar System, the most common types of exoplanets in the galactic planetary census are the so-called Super-Earths/Mini-Neptunes \citep{2013ApJ...766...81F}. These planets have masses a few to a few tens of Earth's mass, are observed very close to their host stars, not rarely in compact multi-planetary configurations \citep{2011arXiv1109.2497M, 2015ARA&A..53..409W,2018AJ....155...48W}. Another relevant characteristic of these type of planets is that, when they appear in multi-planetary systems, their dynamical history is thought to have been shaped by mean motion resonant capture; most of these resonances have subsequently been broken after the disappearance of the disc \citep{2017MNRAS.470.1750I, 2021A&A...650A.152I, 2020MNRAS.494.4950P, 2022Icar..38815206G}, although a few remained stable in resonance and are still observed today (Trappist-1, Kepler-80, Kepler-223 among others). 
Because of their abundance and their importance in planet formation theories, we will concentrate on Super-Earths in this work. Their planet-disc interactions typically fall in the so-called type-I migration regime, where the profile of the gaseous disc around them is only slightly perturbed from the underlying disc structure; the type-II regime instead pertains to those planets that are massive enough to open a noticeable gap around their orbit, thus modifying the gas' structure significantly. We will concentrate here mainly on the type-I regime, although we will also investigate the transition between type-I and type-II migration.

The study of planet-disc interactions in the type-I regime is the subject of a vast number of works, both analytical (e.g.\ \citealt{1979ApJ...233..857G, 1980ApJ...241..425G, 1993ApJ...419..166A, 1997Icar..126..261W, 2008EAS....29..165M}) and numerical (e.g.\ \citealt{2002ApJ...565.1257T, 2004ApJ...602..388T, 2008A&A...487L...9K, 2009A&A...506..971K,  2010MNRAS.401.1950P, 2011MNRAS.410..293P, 2017MNRAS.471.4917J}) Numerical studies resort to hydro-dynamical simulations where a gaseous disc is fully resolved, to investigate the response of the planet to the perturbations driven onto the disc by the presence of the planet itself. This response usually manifests itself as a shrinking of the planetary orbit (a process called inward migration) and a damping of the orbit's eccentricity and inclination (although outward migration and eccentricity excitation can also occur depending on the planet and disc parameters, \citealt{2001A&A...366..263P, 2006A&A...447..369K, 2013A&A...555A.124B, 2006A&A...459L..17P,
2010A&A...523A..30B, 2011A&A...536A..77B}). Although they are extremely important tools, for example when one wants to compare disc models to real observations, such simulations are relatively costly and are for this reason not directly employable in population synthesis works, where instead $N$-body integrations with fictitious forces that mimic planet-disc interactions are preferred (e.g.\ \citealt{2009A&A...501.1139M, 2013A&A...558A.109A, 2017MNRAS.470.1750I,2019A&A...623A..88B,2021A&A...650A.152I, 2020ApJ...892..124O, 2021A&A...656A..69E}).
One of the most widely used prescription for type-I interactions is that of \cite{2008A&A...482..677C}, who fitted dissipative evolution of Super-Earth type planets embedded in a gaseous disc to derive orbital element damping timescales that can be used as efficient recipes in $N$-body integrations.
\cite{2008A&A...482..677C} run 3D simulations, using what are nowadays considered high viscosity values. This may be an issue for low viscosity (or thinner) discs even for lower mass Super-Earth-like planets, as even these may start carving a partial gap. 
Indeed, more recent non-ideal MHD simulations revealed that discs are less viscous than previously thought.
While the question of the effect of different diffusive strengths has been thoroughly investigated for what concerns the planetary migration speed (i.e.\ the torque, e.g.\ \citealt{2011MNRAS.410..293P} which is widely used in the literature in combination with \cite{2008A&A...482..677C}'s prescription for eccentricity and inclination damping; or more recently \citealt{2017MNRAS.471.4917J}), this is not so in the case of eccentricity damping. Besides, as we mentioned earlier, it is known that heavier planets which carve a significant gap may even undergo eccentricity excitation due to planet-disc interaction \cite{2001A&A...366..263P, 2006A&A...447..369K, 2013A&A...555A.124B, 2015ApJ...812...94D}. The question naturally arises on how to properly model the transition from classical (moderately-high-viscosity, thick disc and low-mass planet) type-I eccentricity damping and type-II evolution.

We note that for extremely low viscosities and moderately massive planets, vortices start to appear \citep{2014ApJ...782...88F,2019MNRAS.489L..17M}, making a clear description of planet-disc interactions elusive in these cases. In fact, the analytical formulas for the torque and eccentricity damping found in the literature are for the case of an axisymmetric disc with a smooth profile \citep{2011MNRAS.410..293P, 2017MNRAS.471.4917J}.
Hydrodynamical simulations seem to imply that, in the limit of an inviscid disc, planet-disc interactions for Super-Earths lead to vastly different outcomes than in viscous discs, in that they can hinder capture into resonance and produce completely different systems \citep{2019MNRAS.489L..17M}.
Moreover, analytical formulas that describe planet-disc interactions are typically only valid for eccentricities up to values comparable to the vertical aspect ratio of the disc \citep{2004ApJ...602..388T}.
Due to these limitations, we do not strive to derive general results in an arbitrarily large parameter space, and we instead focus on conditions that are actually normally encountered and useful to the community in a practical sense.\\

In particular, the aim of this paper is to revisit orbital eccentricity damping formulas for levels of viscosity that are more ``modern'' compared to what was used in \cite{2008A&A...482..677C}, just like \cite{2011MNRAS.410..293P} included the effects of viscous diffusion in the expression of the \cite{2010MNRAS.401.1950P} unsaturated torque. The general goal is to better understand the transition between viscous- and inviscid-disc type-I planet-disc interactions. First of all, even low-mass planets that would normally be considered in the type-I regime can, in low viscosity environments, open significant gaps; this can be of interest in the context of the observational techniques mentioned above \citep{2017ApJ...843..127D}. But most importantly, we are specifically interested in the process of resonant capture \citep{2019MNRAS.489L..17M}, since these are the mechanisms that are thought to sculpt the dynamics of the Super-Earth population \citep{2017MNRAS.470.1750I, 2021A&A...650A.152I}. These results will then be directly applicable to exoplanetary population synthesis models.
Since we are not considering inclined orbits here, it is natural to consider a 2D setup. We also note that hydrodynamical investigations of Super-Earths (such as the already mentioned \citealt{2019MNRAS.489L..17M}, but also others such as \citealt{2021A&A...648A..69A}) are run for locally isothermal, 2D discs, since full 3D hydrodynamical simulations for resonant capture are too resource-intensive to be practical.
In order to make a fair comparison with these works, and as a first step in our investigation, we will also consider 2 dimensional, locally isothermal discs in this paper, while future work will be devoted to 3D effects, as well as the expression for inclination damping in partial-gap-opening scenarios. 
Besides, it is well known that resonant capture in the most common resonances (which are mean motion resonance of first order in the eccentricities) does not involve the inclinations, and all known planetary systems in confirmed resonances are observed to be extremely flat (Trappist-1, Kepler-80, Kepler-223, ...). Moreover, 2D-hydro simulations can be made to reproduce the main features of 3D simulations \citep{2009A&A...506..971K, 2012A&A...546A..99K}.

The rest of the paper is organised as follows. In Section \ref{sec:DiscModel} we describe our disc model, while in Section \ref{sec:Methods} we detail our hydrodynamical simulations and the methods used to describe partial gap opening and to calculate orbital element damping timescales, with specific attention to their implementation in $N$-body codes. In Section \ref{sec:Results} we describe the results of our hydrodynamical simulations, while in Section \ref{sec:Discussion} we discuss their implications. In particular, we present a simple recipe to model type-I eccentricity damping that takes into account the effects partial-gap opening and that is consistent with hydrodynamical simulations. Finally, we conclude and summarise our results in Section \ref{sec:Conclusion}. In addition, we give a simple theoretical argument to understand our main conclusions and supplementary explanations on our methods in the Appendix.

\section{Disc Model}\label{sec:DiscModel}
Our hydrodynamical experiments simulate a gaseous 2D disc around a $M_* = M_\odot$ star, extending from 0.35 to 3.3 AU. We assume a power-law profile for the surface density $\Sigma(r) = \Sigma_0 (r/r_0)^{-\alphaSigma}$ and a constant aspect ratio $H/r=h=h_0$ across the disc (flaring index $\betah=0$). We parametrise the turbulent viscosity $\nuvisc$ of the disc using the well-known alpha-prescription \citep{1973A&A....24..337S}: $\nuvisc=\alphaturb H^2 \Omega_\Kepl$; we assume a constant value for $\alphaturb$.
For such a disc, the accretion rate $\dot M_{\gas,\accr}$ is given by $\dot M_{\gas,\accr} = 3\pi\nu\Sigma_\gas = 3 \pi \alphaturb h ^ { 2 } r ^ { 2 } \Omega_\Kepl\Sigma_\gas$. Assuming a constant accretion rate, we set $\alphaSigma = 0.5$.
We also assume that the locally isothermal approximation is valid, that is, we prescribe a temperature profile $T(r) = T_0 (r/r_0)^{-\betaT}$, where $\betaT = 1-2 \betah = 1$ by our aspect ratio prescription. The locally isothermal prescription assumes that cooling timescales are extremely short and is chosen here to provide a fair comparison with other works \citep{2019MNRAS.489L..17M, 2021A&A...648A..69A}. Although more sophisticated thermodynamical assumptions are beyond the scope of this work, we note that they may lead to different outcomes (e.g.\ in terms of gap opening, \citealt{2019ApJ...878L...9M, 2020ApJ...892...65M, 2020MNRAS.493.2287Z}; note however that $e$-damping is similar in isothermal and fully-radiative discs, \citealt{2010A&A...523A..30B}) and will be the subject of future studies.

While the surface density and temperature profiles are fixed, $h$ and $\alphaturb$ are left as free parameters. In our simulations we use $h\in\{0.04, 0.05, 0.06\}$ and $\alphaturb\in\{3.16\times 10^{-5}, 1.\times 10^{-4}, 3.16\times 10^{-4}, 1.\times 10^{-3}\}$.
Inside such a disc, we add a planet at a distance of $a_\pl = 1$ AU using a sinusoidal mass taper to smoothly increase its mass from an initial value of 0 to a final mass of $\mpl/M_*\in \{1 \times 10^{-5}, 3\times 10^{-5},6\times 10^{-5}\}$, corresponding to typical masses of super-Earths, over the course of 50 orbits. The planetary masses are always below the corresponding thermal mass $m_\mathrm{th} = h^3 M_*$, so that the the disc-planet tidal perturbation does not drive local nonlinear shocks and can be treated linearly \citep{1986ApJ...309..846L}. The planet's eccentricity divided by the (fixed) aspect ratio is chosen as $e_\pl/h\in\{0, 0.2, 0.4, 0.6, 0.8, 1\}$. We do not consider higher values for the eccentricity of the planet because the analytical formulas that we wish to compare our results to, start breaking down. In any case, single planets in such mass range have their eccentricities damped by the disc, so extremely large eccentricities are not expected; instead, when multiple planets interact in a disc, e.g.\ by capturing in resonance via convergent migration, the expected capture eccentricities are of order $h$ \citep{2018CeMDA.130...54P}. Thus, considering $e_\pl$ up to a value $\simeq h$ does not result in considerable drawbacks. Along the simulation, the planet's orbit (semi-major axis and eccentricity) is kept fixed.
We thus have a disc and planet setup with three free parameters, namely $h$, $\alpha$, $\mpl$, for each of which we let $e_\pl/h$ vary. Table \ref{tbl:parameters} lists all these parameters with the values chosen in our simulations.

We notice that the system is centered on the star and indirect forces should be considered. Recent works have shown that indirect terms must be carefully taken into account (\citealt{2016MNRAS.458.3918Z}, Crida et al.,\ \emph{in prep}).  The recommendation of Crida et al. is to apply indirect forces to all the elements that feel a direct gravitational force. The planet feels the indirect force due to its own gravity as well as that of the disc. The disc feels indirect forces from the planet. Finally, we do not account for the indirect forces of the disc onto itself since we do not consider the disc's self-gravity. 

\begin{center}
\begin{table}
\centering
\begin{tabular}{ c m{6cm} } 
 \hline
 Parameter & Value  \\ \hline \hline 
 $h$            & $\{0.04, 0.05, 0.06\}$ \\ 
 $\alphaturb$   & $\{3.16\times 10^{-5}, 1.\times 10^{-4}, 3.16\times 10^{-4}, 1.\times 10^{-3}\}$ \\ 
 $\mpl/M_*$         & $\{1 \times 10^{-5}, 3\times 10^{-5},6\times 10^{-5}\}$ \\\hline 
 $e_\pl/h$      & $\{0, 0.2, 0.4, 0.6. 0.8, 1\}$\\
 \hline
\end{tabular}
\caption{Set of free parameters of our simulation. For each value of $h$, $\alphaturb$ and $\mpl$, we run a simulation with $e_\pl/h$ spanning the reported values. The whole set is comprised of 216 high-resolution simulations.}
\label{tbl:parameters}
\end{table}
\end{center}

We implemented this setup in the  fargOCA code (fargo with {\bf C}olatitude {\bf A}dded; \citealt{Lega14})\footnote{The simulations presented in this paper have been obtained with a recently re-factorised version of the code that can be found at: https://gitlab.oca.eu/DISC/fargOCA}. The code is based on the \texttt{fargo} code \citep{2000A&AS..141..165M} extended  to three dimensions. The fluid equations are solved using  a second order upwind scheme with a time-explicit-implicit multi-step procedure.
The code is parallelised using a hybrid combination of MPI and Kokkos \citep{Kokkos1,Kokkos3}. Code units are $G=M_*=1$, and the unit of distance $r_0=1$ is arbitrary when expressed in AU. We use 1024 grid cells with arithmetic spacing in radius (corresponding to a $\delta r\simeq 0.002$) and 3000 cells in azimuth for the full $(0,2\pi)$ (corresponding to a $\delta\phi\simeq 0.002$). Even for the smallest planetary masses, this ensures that we are resolving six cells in a half horseshoe width of such planets, which is needed in order to properly resolving the co-rotation torque \citep{2011MNRAS.410..293P, Lega14}. We checked the convergence of our results by halving and doubling the resolution, seeing no discernible difference in the outcome.
We used a smoothing length for the potential of the planet of $r_\mathrm{sm} = \epsilon H(r)$ with $\epsilon = 0.6$, which better reproduces 3D effects in 2D simulations \citep{2012A&A...541A.123M}; finally, we used evanescent boundary conditions \citep{ValBorro2006}.

\section{Methods}\label{sec:Methods}
\subsection{Disc density profile for partial-gap opening planets}\label{subsec:PGO}
A planet orbiting a star inside its protoplanetary disc will affect the disc structure in different ways depending on the system's parameter. If the planet's mass is low, its gravitational effect onto the disc is small enough that the disc structure does not change significantly from the background unperturbed disc profile. This is the so-called type-I regime of planet-disc interactions. More massive planets will instead exert a torque onto the disc that can overcome the restoring viscous torque from inside the gas itself, and the planet will carve a gap around its orbit. This is the so-called type-II regime of gap-opening planets.
For MRI-type levels of viscosities ($\alphaturb \gtrsim 10^{-3}$), the type-I regime usually holds for planets up to a few tens of Earth masses.
However, for sufficiently thin discs and at sufficiently low viscosities, (for example in the MRI-dead zone, e.g.\ \citealt{2014prpl.conf..411T} for a review), even a small planet may start opening a partial gap.

We thus investigate the depth of the gap carved by simulated planets in a disc, to establish a threshold down to which type-I type interactions can be considered valid. 
\cite{2006Icar..181..587C} described the gap opened by a planet on a circular orbit by balancing gravity, viscous and pressure torques (the latter ones originating from the evacuation by pressure supported waves of gravitational torques). They define (arbitrarily) that a planet has opened a gap when the equilibrium disc surface density is 10\% of the unperturbed disc surface density, $\Sigma_{\min}/\Sigma_0=0.1$, and derive a condition for gap opening given by
\begin{equation}\label{eq:Crida2006GO}
    \frac{3}{4} \frac{H}{R_\mathrm{H}}+\frac{50}{q \mathcal{R}}\lesssim 1,
\end{equation}
where $R_\mathrm{H}$ is the Hill radius of the planet, $q=\mpl/M_*$ and $\mathcal{R} = r_\pl^2 \Omega_{\Kepl,\pl}/\nuvisc$ is the Reynolds number.

\citep{2018ApJ...861..140K} gave a prediction for the value of $\Sigma_{\min}/\Sigma_0$, i.e.\ of the gap depth produced by a planet on a circular orbit, depending on the physical parameters of the disc and the planet. They showed that
\begin{equation}\label{eq:KanagawaGapDepth}
    \Sigma_{\min}/\Sigma_0\simeq \frac{1}{1+0.04 K}
\end{equation}
where
\begin{equation}\label{eq:KanagawaK}
    K: = q^2 h^{-5} \alphaturb^{-1},
\end{equation}
is a dimensionless parameter. \cite{2019ApJ...884..142G} slightly improved this result, replacing \eqref{eq:KanagawaGapDepth} with $\Sigma_{\min}/\Sigma_0\simeq 1/(1+0.046 K)$. Finally, \cite{2020A&A...643A.133B} considered the effect of gas accretion in shaping the gap profile and gap opening mass for giant planets (typically $q\gtrsim 10^{-4}$); since we are again interested in type-I interactions, we do not consider significant gas accretion in this work, and instead keep our planetary masses fixed in time after the initial mass taper ramp-up.

It is interesting to note that \citep{2018ApJ...861..140K} show that the transition between type-I and type-II migration happens at values of $K$ of a few tens, which corresponds to a gap of order 0.5; their results seem however to depend on the level of viscosity, with the transition happening at larger values of $K$ for lower viscosities. Instead, \cite{2006Icar..181..587C} again considers a gap opened (and thus transition into type-II) when the gap is 10\%. We can thus consider for our purposes that the transition from type-I to type-II migration happens when the gap depth is of a few $10^{-1}$, and we won't investigate the evolution below this threshold.

Since the eccentricities are small, we consider, for each setup shown in Table \ref{tbl:parameters}, the circular case as the nominal case to derive the gap profile. This allows us to compare our results with the results from the literature. 
We run all our $e_\pl=0$ simulations up to $t_\mathrm{max} = $ 3000 orbits of the planets, as was done in \citealt{2018ApJ...861..140K}. In the lowest viscosity case, we integrated for an additional 1000 orbits to ensure the reaching of a steady-state surface density profile. Indeed, we check that over the last 50 orbits of the planet, the surface density does not change by more than $\sim 0.1$ -- $0.5\%$ (with the longest time needed for convergence given by the lowest-viscosity cases); as a further test of convergence, we integrated 4 setups spanning all the different $\alphaturb$ values for an additional 3000 orbits and checked that the relative difference in surface density over the last additional 3000 orbits is less than $\simeq 2\%$.
We then consider the surface density contrast $\Sigma_{t_\mathrm{max}}(r)/\Sigma_0(r)$, where $\Sigma_t$ is the (azimuthally averaged) surface density of the disc at time $t$, as a function of the radial distance $r$. 
We mark the minimum value $\min_{r}\left[\left(\Sigma_{t_\mathrm{max}}/\Sigma_0\right)(r)\right]$ of the surface density contrast at time $t_\mathrm{max}$ and define it as the gap depth (or take an average of the two minima found slightly exterior and interior to the planet's orbit), which we denote with $\Sigma_{\min}/\Sigma_0$. Our results are presented in section \ref{subsec:PartialGapResult}.

\subsection{Orbital elements damping timescales}\label{subsec:OEDT}
Along the simulation, at constant time-intervals, the \texttt{fargOCA} code outputs two sets of forces: the (direct) force felt by the planet on its fixed orbit from the disc and the force felt by the star from the disc. The second force would not have any effect if the disc were axisymmetric, but because of the gas' response to the presence of the planet, the induced asymmetry causes a net force felt by the star. Since in our simulation the frame of reference is astrocentric, we are not in an inertial reference frame. This means that the force felt by the star from the disc will result in an indirect (fictitious) response force felt by the planet in the astrocentric reference frame. Thus, we need to add this force to the one describing the direct planet-disc interaction. This yields a force $\bfF_{\disk\to\pl}$ which describes the sum of direct and indirect planet-disk interactions, and thus the true force felt by the planet in an inertial reference frame.

From this force, we obtain the torque
\begin{equation}
    \bfGamma := \bfr_\pl \times \bfF_{\disk\to\pl},
\end{equation}
and the power
\begin{equation}
    P := \bfF_{\disk\to\pl} \cdotp \bfv_\pl.
\end{equation}
Note that, since we are on the plane, the vector $\bfGamma = (0,0,\Gamma_z)^\intercal$, so we can concentrate on the scalar quantity $\Gamma := \Gamma_z = \|\bfGamma\|$.
For a planet on an eccentric (planar) orbit, both the torque $\Gamma$ and the power $P$ are needed in order to determine the response of its orbital elements to the force $\bfF_{\disk\to\pl}$ (we use orbit-averaged forces, where the average is done over 20 points along the planet's eccentric orbit). This can be easily done as follows.

By definition, $\Gamma := \frac{\D\ANGMOM_\pl}{\D t} = \dot{\ANGMOM_\pl}$, the rate of change of the (orbital) angular momentum
\begin{equation}
    \ANGMOM_\pl := \redmass \sqrt{\GM a (1-e^2)},
\end{equation}
where $\redmass = (m_\pl M_*)/(M_*+m_\pl) \approx m_\pl$ is the reduced mass of the planet, $\GM = \GravC (M_*+m_\pl) \approx \GravC M_*$ is the reduced gravitational parameter, and $a$ and $e$ are the semi-major axis and eccentricity of the planet's orbit. At the same time, by definition $P := \frac{\D E_\pl}{\D t} = \dot{E_\pl}$, the rate of change of the (orbital) energy
\begin{equation}
    E_\pl := -\frac{\GM \redmass}{2a}.
\end{equation}
By the fundamental equation $\Gamma = \dot{\ANGMOM_\pl}$, the angular momentum of the planet evolves, and we can introduce a migration timescale $\tau_\mig$ defined as
\begin{equation}\label{eq:taumigDEF}
    \frac{\dot{\ANGMOM_\pl}}{\ANGMOM_\pl} =: -\frac{1}{\tau_\mig};
\end{equation}
this is however not the semi-major axis evolution timescale, and in the case of eccentric orbits the evolution of the eccentricity must also be taken into account. Indeed, we also define 
\begin{align}
\label{eq:tauaDEF}
    \frac{\dot{a}}{a}   &=: -\frac{1}{\tau_a},\\
\label{eq:taueDEF}
    \frac{\dot{e}}{e}   &=: -\frac{1}{\tau_e},
\end{align}
in order to represent the evolution timescales of the semi-major axis and of the eccentricity of the planet. We now need to express these timescales in terms of the torque, power, angular momentum and energy of the planet.

Taking the time derivative of the angular momentum and the energy, we obtain:
\begin{align}
    P       &= \dot{E_\pl} = \frac{\GM \redmass}{2a^2} \dot{a} = E_\pl \left(-\frac{\dot{a}}{a}\right) \equiv \frac{E_\pl}{\tau_a}\\
    \Gamma  &= \dot{\ANGMOM_\pl} = \redmass \sqrt{\GM} {\left(\frac{1}{2}a^{-1/2} \dot a (1-e^2)^{1/2} + \frac{1}{2}a^{1/2}(1-e^2)^{-1/2}(-2 e \dot{e})\right)}\nonumber \\
    &= \ANGMOM \left(\frac{1}{2}\left(\frac{\dot{a}}{a}\right) - C(e) \left(\frac{\dot{e}}{e}\right)\right) \equiv \ANGMOM \left(-\frac{1}{2\tau_a}+\frac{C(e)}{\tau_e}\right),
\end{align}
with 
\begin{equation}
C(e) = \frac{e^2}{1-e^2},
\end{equation}
$C(e)\approx e^2$ for small eccentricities. Thus we have
\begin{align}
    \tau_\mig   &= -\frac{\ANGMOM_\pl}{\Gamma},\\
    \tau_a      &= \frac{E_\pl}{P},\\
\label{eq:taue.wrt.taumig-taua}
    \tau_e      &= C(e) \left(-\frac{1}{\tau_\mig} + \frac{1}{2\tau_a}\right)^{-1}.
\end{align}
We also see that for a circular orbit $\tau_\mig = 2\tau_a$.

\begin{figure}[ht!]
\centering
\includegraphics[width= 0.5\textwidth ]{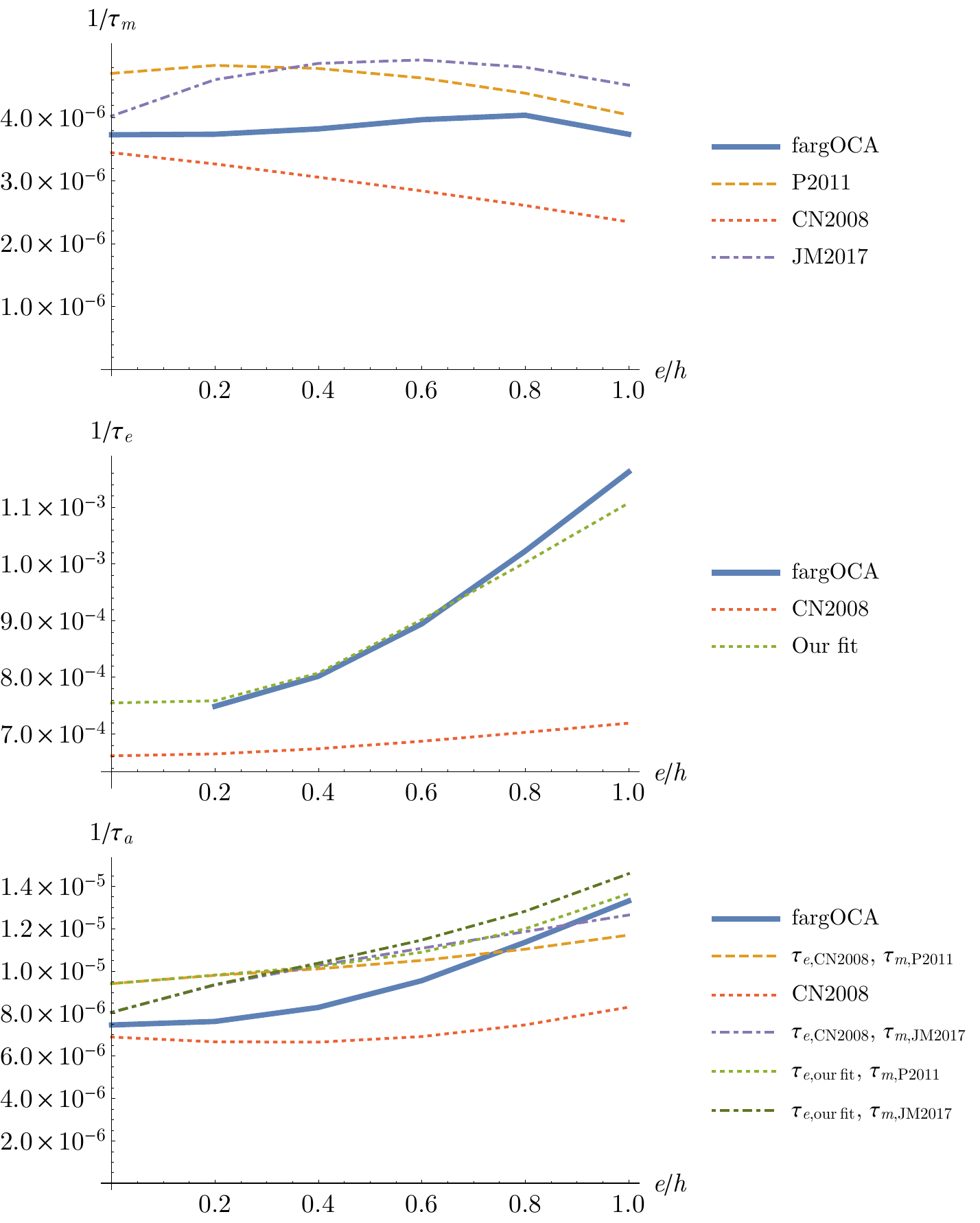}
\caption{Comparison of the orbital elements damping efficiencies $1/\tau_\mathrm{mig}$ (upper panel), $1/\tau_e$ (middle panel) and $1/\tau_a$ (lower panel) obtained from the output of our hydro simulations and from analytical formulas in the literature (CN2008 for \citealt{2008A&A...482..677C}, P2011 for \citealt{2010MNRAS.401.1950P}, and JM2017 for \cite{2017MNRAS.471.4917J}). The figures here are for the case of a $m_\mathrm{pl}=1\times 10^{-5} M_*$ planet, with $\alpha = 3\times 10^{-4}$ and $h=0.05$.  $\tau_\mathrm{mig}$ is obtained from the torque felt by the planet from the disc, and typically gives good agreement between hydro and analytical predictions within the latter's margin of uncertainty (20\%). The analytical expression for $\tau_e$ comes from \citealt{2008A&A...482..677C} or from our fit presented in Sect.\ \ref{subsec:eDampEffResult}. $\tau_a$ is obtained from $\tau_\mathrm{mig}$ and $\tau_e$ using Eq.\ \eqref{eq:taua.wrt.taumig-taue}.}
\label{fig:ComparisonPlot}%
\end{figure}

\subsection{Planet-disc interactions in $N$-body codes}
$N$-body codes implement type-I migration using the timescales $\tau_\mig$ and $\tau_e$ to define accelerations onto the planet given by (Papaloizou \& Larwood 2000)
\begin{align}
    \bfa_\mig   &= -\frac{\bfv_\pl}{\tau_\mig},\\
    \label{eq:tau_e.PL2000}
    \bfa_e      &= -2\frac{(\bfv_\pl\cdotp\bfr_\pl)\bfr_\pl}{r_\pl^2 \tau_e},
\end{align}
where $\bfr_\pl$ and $\bfv_\pl$ are the planet's position and velocity.
The first equation describes a change in angular momentum, that is a torque, so that the angular momentum evolves according to \eqref{eq:taumigDEF}; the second equation represents a force that has zero torque since it is a radial force, so it does not contribute to $\Gamma$ but only to the power, and implements, for small $e$'s,\footnote{
By applying \eqref{eq:tau_e.PL2000} over an orbit, the quantity that is damped exponentially (over a timescale $\tau_e/2$) is $E=(1-e^2)^{-1/2}-1\simeq e^2/2$ for small $e$'s. The quantity $E$ is the ratio between the AMD of the planet \citep{1997A&A...317L..75L} and the norm of the angular momentum vector.} an orbital damping of the eccentricity as described by equation \eqref{eq:taueDEF}. The semi-major axis evolution described by \eqref{eq:tauaDEF} thus results from a combination of torque and $e$-damping with a timescale given by 
\begin{equation}\label{eq:taua.wrt.taumig-taue}
\tau_a = \left(\frac{1}{\tau_\mig/2} +  \frac{C(e)}{\tau_e/2}\right)^{-1},
\end{equation}
that is, the equivalent of \eqref{eq:taue.wrt.taumig-taua}.\\

\subsubsection{Type-I forces}
Various formulas exist in the literature to implement fictitious type-I forces in this framework. 
\cite{2008A&A...482..677C} give explicit expressions for $\tau_\mig$ and $\tau_e$ (and $\tau_i$, the damping of the inclinations for non-planar orbits, not considered in this paper) by fitting the orbital evolution of a $m_\pl=10\mEarth$ planet placed on a variety of initial configurations. They used 3D simulations with a relatively low resolution (less than 2 cell per half horseshoe width of the planet, while in order to resolve the corotation torque one needs at least six cells, \citealt{2011MNRAS.410..293P}) and a fixed viscosity of $\alpha=5\times10^{-3}$. Defining the typical type-I damping timescale $\tau_\mathrm{wave}$ \citep{2004ApJ...602..388T},
\begin{equation}
    \tau_\mathrm{wave} = \left(\frac{M_{\odot}}{m_\mathrm{pl}}\right) \left(\frac{M_\odot}{\Sigma_\mathrm{gas,pl} a_\mathrm{pl}^2}\right) h^4 \Omega_{K,\mathrm{pl}}^{-1},\label{eq:tauwave}
\end{equation}
their best fit yielded (in the planar case)
\begin{equation}
    \tau_{\mig, \mathrm{CN2008}} =\frac{2 \tau_\mathrm{wave}}{2.7 + 1.1\alphaSigma} h^{-2}\times P_e,
\end{equation}
where
\begin{equation}\label{eq:P_e}
    P_e = \frac{1+(\hat e/2.25)^{1.2} + (\hat e/2.84)^6}{1-(\hat e/2.02)^4}
\end{equation}
is a reduction factor due to the planet's eccentricity and $\hat e = e/h$, while
\begin{equation}\label{eq:taue_CN2008}
    \tau_{e, \mathrm{CN2008}} = \frac{\tau_\mathrm{wave}}{0.780}\left[1-0.14\left(\frac{e}{H/r}\right)^2+0.06\left(\frac{e}{H/r}\right)^3\right].
\end{equation}

\citealt{2010MNRAS.401.1950P} focused instead on the explicit expression for the total torque experienced by a planet on a circular orbit, coupling linear estimates of the Lindblad torque $\Gamma_\Lind$ to a non-linear model of corotation torques. The latter torques are split into a barotropic (or vortensity-driven) component and an entropic (or thermal) component. Unlike the Lindblad torques, corotation torques are prone to saturation, so that only the Lindblad torque would remain without any restoring process. Thus, \cite{2011MNRAS.410..293P}, studied the effects of diffusion in restoring the corotation torque $\Gamma_\Corot$, whereby in the limit of very strong diffusion, the linear corotation torque can be recovered. Their formulas are thus viscosity-dependent and for this reason are more cumbersome than those of \cite{2008A&A...482..677C}, so we don't re-write them here. They capture the behaviour of the torque observed in their 2D numerical experiments with an error of up to 20\%. 
\cite{2017MNRAS.471.4917J} re-analysed the torque experienced by low- and intermediate-mass planets using 3D simulations, deriving improved formulas valid for planets that do not significantly deplete their coorbital region. In addition to the vortensity and entropic components (scaling respectively with the gradient of vortensity and entropy) they consider a temperature component (scaling with the temperature gradient) and a so-called viscous coupling term. By striving for accuracy rather then simplicity and splitting the corotation torque into different components, \cite{2017MNRAS.471.4917J}'s formulas are significantly more cumbersome, and we again do not reproduce it here. Note however than in our locally isothermal case, only the vortensity and temperature components appear.\\

An equivalent of the aforementioned studies on the detailed characterisation of the viscosity-dependent torque onto a planet but in the case of an eccentric planet (i.e. a formula for the power including diffusive effects) has not yet been carried out.
Still, population synthesis models need to include both $a$- and $e$-damping (and inclination damping, not considered in this work), since planet-planet interaction can excite the planets' eccentricities and thus the circular approximation is not adequate in a practical sense. Many works \citep{2017MNRAS.470.1750I, 2021A&A...650A.152I, 2021A&A...656A..69E} have implemented instead a {\it mélange} of \cite{2011MNRAS.410..293P}'s formula for the torque plus \cite{2008A&A...482..677C}'s expression for the eccentricity damping, eq. \eqref{eq:taue_CN2008}. Moreover, the torque itself has to be modified to include an $e$-dependency \cite{2010A&A...523A..30B, 2011A&A...536A..77B, 2013A&A...553L...2C, 2013A&A...558A.105P, 2014MNRAS.437...96F}: this dependency introduces a reduction of the Lindblad torque like the one found by \cite{2008A&A...482..677C},
\begin{equation}
    \Delta_\Lind = P_e^{-1},
\end{equation}
and a reduction of the corotation torque given by
\begin{equation}
    \Delta_\Corot = \exp{\left(-e/e_\mathrm{f}\right)},
\end{equation}
where $e_\mathrm{f}=0.5 h + 0.01$ is defined in \citealt{2014MNRAS.437...96F}. The total torque is thus
\begin{equation}
    \Gamma_\tot = \Delta_\Lind \Gamma_\Lind + \Delta_\Corot \Gamma_\Corot.
\end{equation}

Figure \ref{fig:ComparisonPlot} contains an example of the comparison of orbital damping timescales, from one of our hydrodynamical simulations and from the formulas from the literature (CN2008 for \citealt{2008A&A...482..677C}, P2011 for \citealt{2010MNRAS.401.1950P}, and JM2017 for \cite{2017MNRAS.471.4917J}). The migration timescale, and thus the torque, gives good agreements with these analytical predictions within their typical errors (of the order of 20\%). The $e$-damping timescale is instead more efficient in the case shown here compared to the prediction from \citealt{2008A&A...482..677C}. We will compile our results for all of our numerical sample in section \ref{subsec:eDampEffResult}.
 
\subsubsection{Transition into type-II migration regime}
Although we will not deal with the type-II migration regime in this paper, it is instructive to see how the transition from type-I to type-II migration is usually handled in $N$-body codes.
Indeed, even though it is conceptually easy to think in terms of separate migration regimes, recent developments in our understanding of disc structure and evolution have shown that the processes that drive turbulent viscosities, such as the MRI, might be quenched in large portions of the midplanes where planets form; the remaining hydro-instability-driven viscosities would therefore be much lower than expected, so that even a low-mass planet may start opening a partial gap.

The hydrodynamical simulations of \citealt{2018ApJ...861..140K} allowed them to express the timescale of type-II migration $\tau_{\mathrm{mig, II}}$ to the type-I migration timescales modulated by the gap depth $\Sigma_{\min}/\Sigma_0$
\begin{equation}\label{eq:type-ItoIIMigTransition}
    \tau_{\mig,\mathrm{II}} = \left(\frac{\Sigma_{\min}}{\Sigma_0}\right)^{-1} \tau_{\mig,\mathrm{I}}.
\end{equation}
This can be used to transition from the two regimes in the case of partial-gap opening planets. After the gap is considered to have been fully opened (e.g.\ from \cite{2006Icar..181..587C} condition \eqref{eq:Crida2006GO}), the type-II migration regime \citep{1986ApJ...309..846L, 2018A&A...617A..98R} can be considered fully operational.
We stress again that, so far, for the type-I regime only the migration timescale is modulated by the opening of a partial gap in population synthesis works, but not the eccentricity damping. Thus, in section \ref{subsec:eDampEffResult} we will attempt at obtaining an equivalent adaptation on \eqref{eq:type-ItoIIMigTransition} for the eccentricity damping for partial-gap opening planets.

\section{Results}\label{sec:Results}

\subsection{Emergence of vortices}
Of our setups listed in Table \ref{tbl:parameters}, not all lead to a stable steady-state disc profile, as some of them showed the emergence of vortices. 
We choose not to include in the final analysis presented in the next sections those setups where vortices occurred (we also tested that vortices appear in these setups even when a smoother mass taper is used, namely a ramp-up time longer by a factor 10). This is because, even though vortices might dissipate over time (this typically happens over a few hundreds to thousands of orbits in our simulations), such cases cannot be fully captured by a simple analytical expression for type-I planet-disc interactions. In any case, we will see that vortices appear in our simulations when gaps are significantly deep for migration to be considered of the type-II regime, which is beyond the scope of this work. We note that whenever the $e_\pl = 0$ simulation did not show the emergence of a vortex, so did the $e_\pl\neq 0$ runs with the same value of $h$, $\alphaturb$ and $\mpl$. This is advantageous both in theory and in practice: our results on the emergence of a vortex do not depend on the planetary eccentricity (across the values considered in this work), and each valid setup without vortices will contribute a data point for each value of the eccentricity. 
Five setups, 
($h=0.04$, $\alphaturb=10^{-4}$, $\mpl/M_*=6\times 10^{-5}$),
($h=0.04$, $\alphaturb=3.16\times10^{-5}$, $\mpl/M_*=6\times 10^{-5}$),
($h=0.04$, $\alphaturb=10^{-4}$, $\mpl/M_*=6\times 10^{-5}$),
($h=0.05$, $\alphaturb=3.16\times10^{-5}$, $\mpl/M_*=3\times 10^{-5}$) and 
($h=0.05$, $\alphaturb=3.16\times10^{-5}$, $\mpl/M_*=6\times 10^{-5}$),
showed the emergence of a vortex. These correspond, as a rule of thumb, to the cases of a lower $\alphaturb$ or $h$ value, or highest $\mpl$, as expected in a qualitative sense.  
These simulations allow us to obtain a quantitative constraint on the system's parameters which would lead to the emergence of a vortex. We give this constraint in the next section in terms of the depth of the gap carved by the planet. 
The case ($h=0.05$, $\alphaturb=10^{-3}$, $\mpl/M_*=1\times 10^{-5}$), where a vortex does not appear, is shown as an example in Figures \ref{fig:NoVortex} and \ref{fig:NoVortex_Torque} in Appendix \ref{apx:TypicalDiscEvo}.

\subsection{Partial gap opening at low viscosity and/or thin discs}\label{subsec:PartialGapResult}
\begin{figure}[t!]
\centering
\includegraphics[width= 0.45\textwidth ]{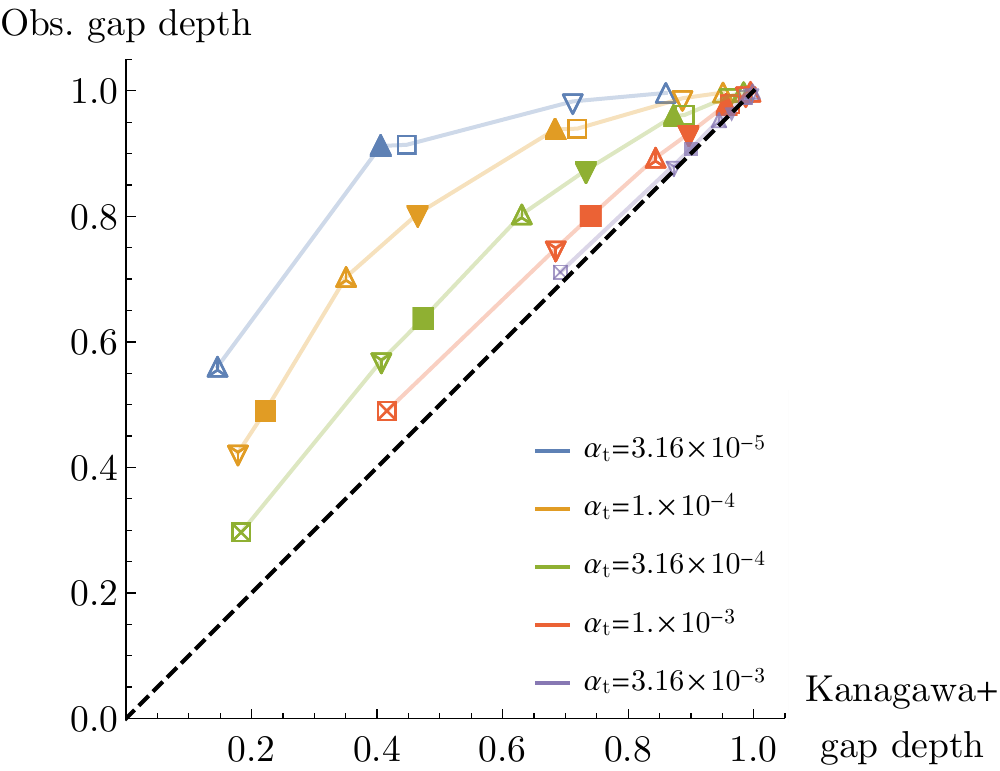}
\caption{Predicted gap depth $\Sigma_{\min}/\Sigma_0$ from \citealt{2018ApJ...861..140K} vs.\ value obtained from the simulations. Different colours represent different levels of viscosity as shown in the legend. The lower the viscosity, the bigger is the difference in the gap depth. However, down to the viscosities considered by \citealt{2018ApJ...861..140K}, we see good agreement as expected (the data for $\alphaturb=3.16\times 10^{-3}$ are shown with smaller symbols as they are only used for the analysis of the gap depth). Symbols of different shapes are used to represent different aspect ratios and planetary masses: squares, downward-pointing triangles and upward-pointing triangles represent aspect ratios of 0.04, 0.05 and 0.06 respectively; empty, filled and crossed symbols represent $\mpl/M_*=10^{-5}$, $3\times 10^{-5}$ and $6\times 10^{-5}$ respectively.}
\label{fig:KanagawaVsObs_GapDepth}%
\end{figure}

\begin{figure}[t!]
\centering
\includegraphics[width= 0.45\textwidth ]{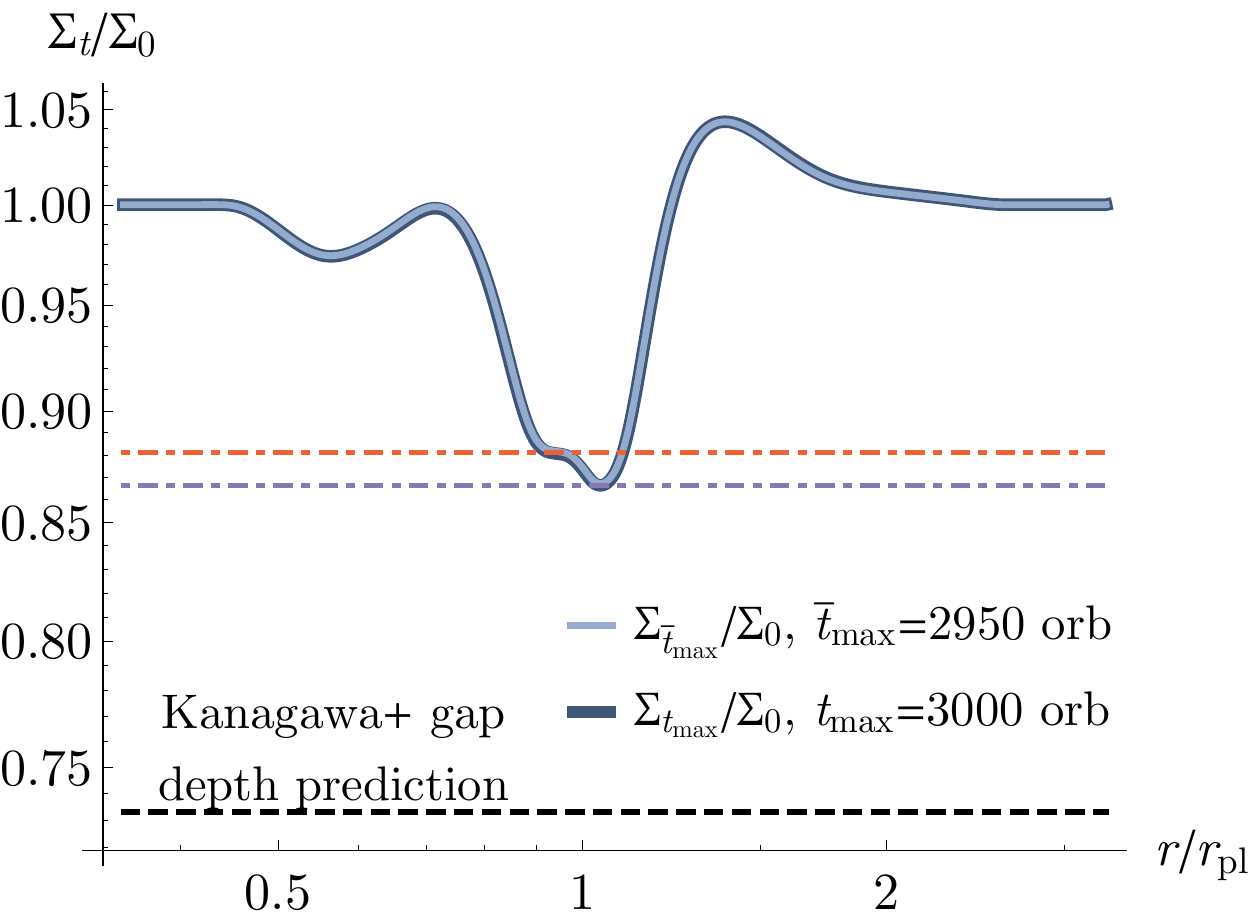}
\caption{Surface density profile $\Sigma_t/\Sigma_0$ after $\bar{t}_{\max}=2950$ orbits and $t_{\max}=3000$ orbits of a $\mpl/M_* = 3\times 10^{-5}$ planet in a disc with $\alphaturb=3.16\times 10^{-4}$, and $h=0.05$. Note that after the last 50 orbits the surface density has reached a steady state, with a change of less than 0.1\%. The black horizontal dashed line represents the prediction of the gap depth from \citep{2018ApJ...861..140K}. The two dot-dashed horizontal lines are used here to measure the observed depth of the gap carved by the planet by taking the average of their values.}
\label{fig:SurfDens_mpl-3.e-5_alpha-3.16e-4_h-0.05}%
\end{figure}

We calculated the gap depth $\Sigma_{\min}/\Sigma_0$ after $t_\mathrm{max} = 3000$ orbits of the planet (or for an additional 1000 orbit for the lowers viscosity cases), when a steady-state was reached, as described in Section \ref{subsec:PGO}.
Figure \ref{fig:KanagawaVsObs_GapDepth} shows the outcome of our simulations compared to the predicted gap depths from \citealt{2018ApJ...861..140K}'s formula \eqref{eq:KanagawaGapDepth}. We note that they considered viscosities down to $\alphaturb=10^{-3}$, while we go as low as $\alphaturb=3.16\times 10^{-5}$. In order to make a more fair comparison, we run additional hydro-simulations with the same $\mpl$ and $h$ ranges from Table \ref{tbl:parameters} for $\alphaturb=3.16\times 10^{-3}$, in the $e=0$ case alone, with the sole purpose of checking our gap-depth results with \citealt{2018ApJ...861..140K}'s. We see that for $\alphaturb=1$ -- $3\times 10^{-3}$ our experiments give a good match, as expected, with the results from \citealt{2018ApJ...861..140K}, well within the spread observed in their simulations around their fit. However, for lower and lower viscosities, we observe a significant difference between the predicted value and the observed one. In general, \citealt{2018ApJ...861..140K}'s formula \eqref{eq:KanagawaGapDepth} over-estimates how deep a gap a given planet will carve. We show in figure \ref{fig:SurfDens_mpl-3.e-5_alpha-3.16e-4_h-0.05} an example with $\mpl/m_* = 3\times 10^{-5}$, $\alphaturb=3.16\times 10^{-4}$, and $h=0.05$: the observed gap depth after 3000 orbits is $\simeq 0.9$, while the predicted value is $\simeq 0.7$. Note that after 3000 orbits of the planet, the surface density changes by less than 0.1\% over 50 orbits.

\begin{figure}[t!]
\centering
\includegraphics[width= 0.45\textwidth ]{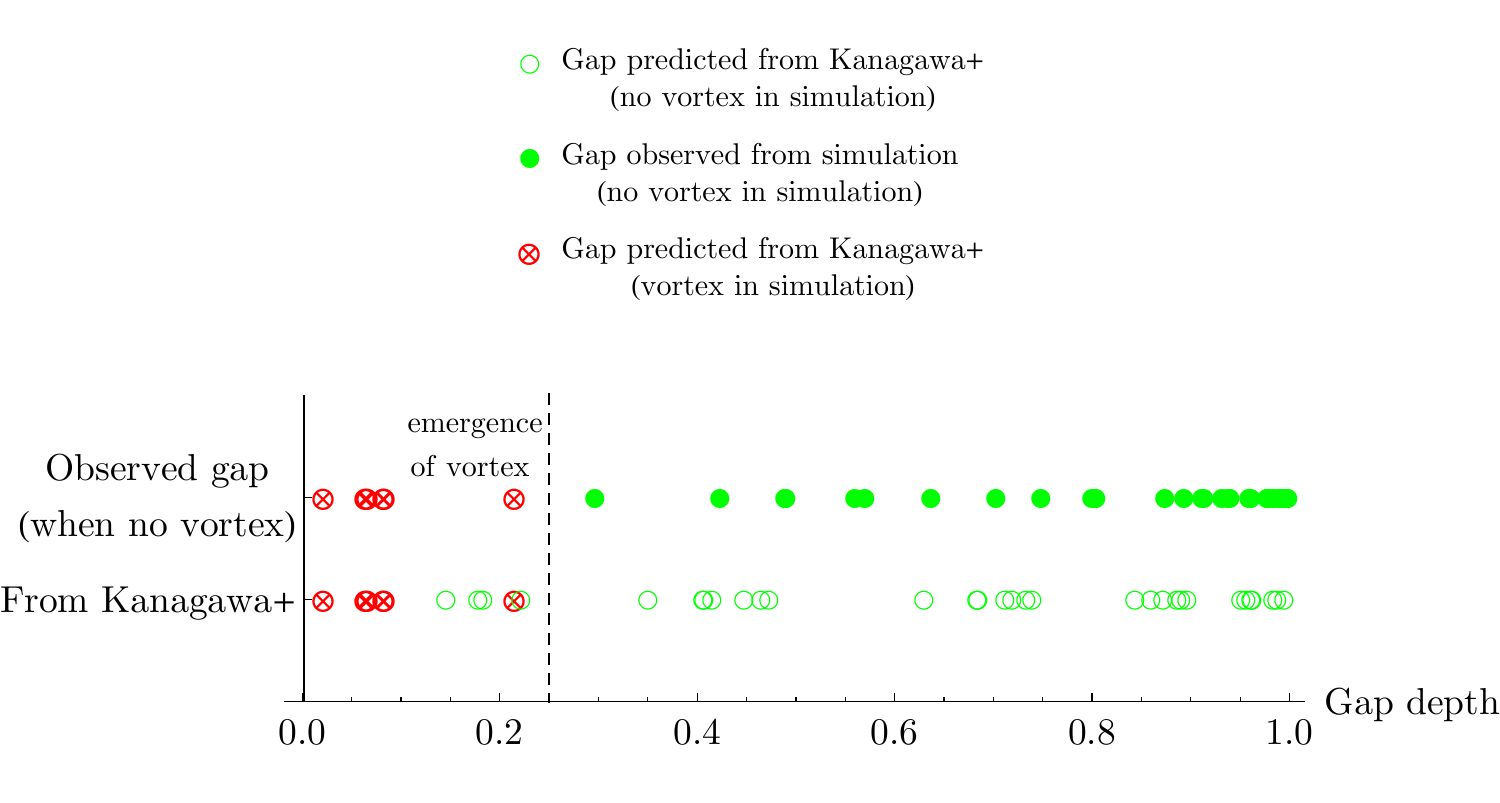}
\caption{Emergence of vortices in our simulations, as a function of the gap depth. When no vortex was observed, we report both the observed gap depth at the end of our simulations, as well as the predicted one from \citealt{2018ApJ...861..140K}, with a green circle (filled and unfilled respectively). A red $\otimes$ symbolises those simulations where a vortex appeared, in which case only the \citealt{2018ApJ...861..140K}'s predicted value is plotted. A dashed vertical line marks the approximate location where the transition between no vortex and vortex lies.}
\label{fig:VortexLowerLimitPlot}%
\end{figure}

Another feature seen in Figure \ref{fig:KanagawaVsObs_GapDepth} is that the lower-viscosity cases are represented by fewer points. This is because in some of the simulations (namely those with more massive planet and/or thinner discs) a vortex has appeared which does not allow us to draw any definitive conclusion. Based on our numerical experiments, we can describe the limit in the parameter space after which one can expect to see a vortex in such simulations as a function of the gap depth. Figure \ref{fig:VortexLowerLimitPlot} shows a diagram where we label the outcome of each simulation green when a vortex has not appeared and red when it has appeared. When a vortex has not appeared, we report both the observed gap depth (filled circle) and the predicted gap depth (unfilled circle) from \citep{2018ApJ...861..140K}; when a vortex has appeared, we cannot use the simulations to observe a gap depth, and thus we only report the predicted gap depth from \citep{2018ApJ...861..140K}. We see that, in both cases, vortices are expected to appear when the gap depth $\Sigma_{\min}/\Sigma_0\lesssim 0.25$.

\subsection{Eccentricity damping efficiency for partial-gap opening planets}\label{subsec:eDampEffResult}
\begin{figure*}[ht!]
   \resizebox{\hsize}{!}
        {
            \subfigure[]{
            \includegraphics[width=0.45\textwidth]{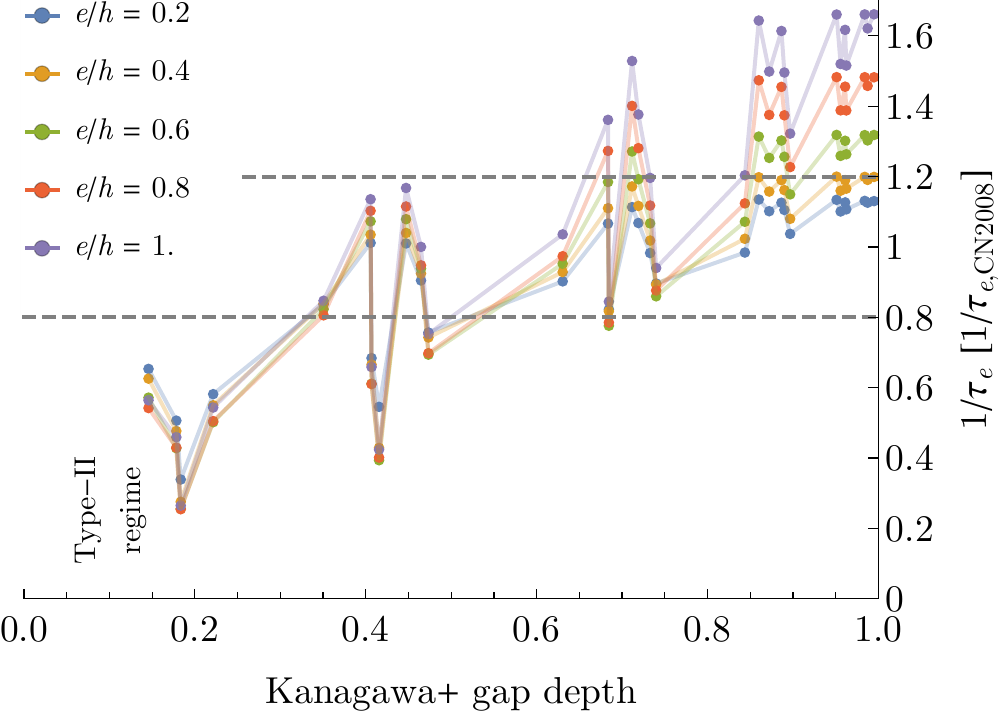}
            }%
            \subfigure[]{
            \includegraphics[width=0.45\textwidth]{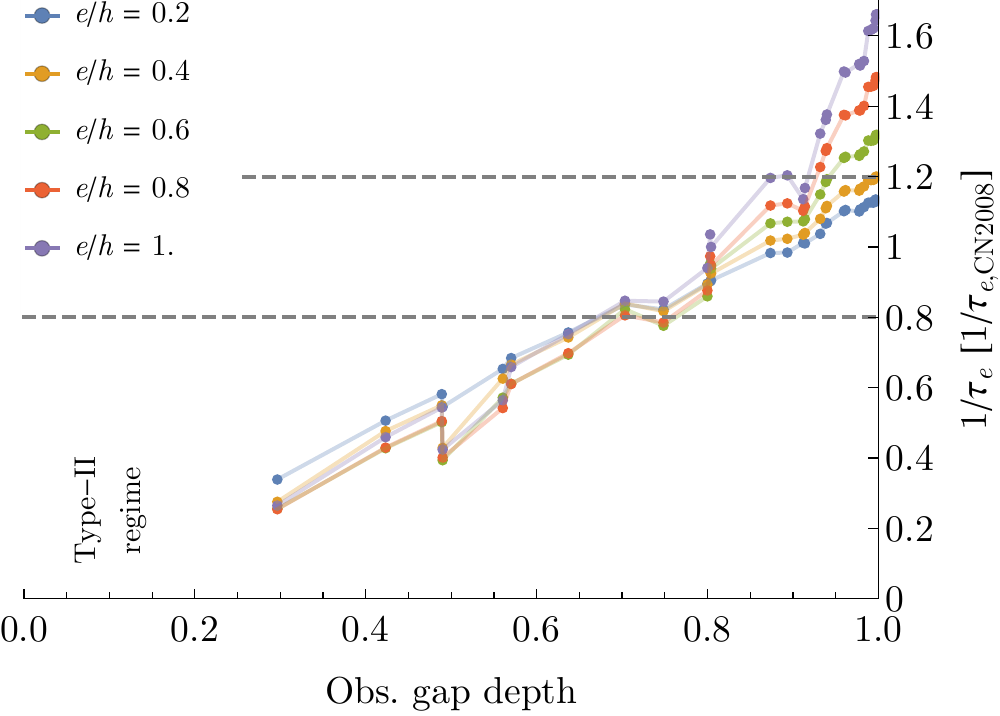}
            }%
        }
      \caption{Eccentricity damping efficiency versus gap-depth for all setups where no vortex emerged. In both panels, the $e$-damping efficiency on the vertical axis is normalised by the expected value from \cite{2008A&A...482..677C}. A 20\% error around the expected value in the limit of no gap (to the right in the plots) is shown by two dashed horizontal gray lines. Values for the $e$-damping efficiency are shown for different eccentricities $e/h \in \{0.2, 0.4, 0.6, 0.8, 1\}$ by points of different colours joined together by opaque lines. Panel (a) shows on the horizontal axis the gap depth according to \cite{2018ApJ...861..140K}'s prediction, which results in an extremely noisy scatter plot. 
      Panel (b) shows on the horizontal axis the gap depth observed from the actual simulations as plotted in Figure \ref{fig:KanagawaVsObs_GapDepth}, which shows a much cleaner dependence. In both panels, the overall trend clearly shows a decrease in $e$-damping efficiency (i.e.\ longer $e$-damping timescales) for deeper and deeper gaps, down to a factor of $\sim 1/4$ less efficient eccentricity damping at the transition from type-I to type-II regimes (gap depths of $\simeq 0.25$) as compared to the limit of no gap.
      }
\label{fig:edampVsGapDepth}
\end{figure*}

Following section \ref{subsec:OEDT}, we extract the eccentricity damping efficiency $1/\tau_e$ from our hydrodynamical simulations for all setups where no vortex has emerged. Figure \ref{fig:edampVsGapDepth} shows the eccentricity damping efficiency $1/\tau_e$ normalised by the expected efficiency $1/\tau_{e, \mathrm{CN2008}}$ from \cite{2008A&A...482..677C} (equation \eqref{eq:taue_CN2008}), which is the one that has been used extensively in the literature so far (e.g.\ \citealt{2017MNRAS.470.1750I, 2021A&A...650A.152I, 2021A&A...656A..69E}). We show the damping efficiency as a function of the gap depth carved by the planet (see previous sections), both using \cite{2018ApJ...861..140K}'s prediction (panel a) and the gap obtained from each hydrodynamical simulation in the circular case (panel b). While panel (a) yields a very noisy plot, it is clear in panel (b) that there is a strong trend when one uses the actual gap depth. 

The overall trend is however clear in both plots: the eccentricity damping is less efficient for deeper gaps (left side of the plots), down to a factor of $\sim 1/4$ less efficient at gap depths of $\simeq 0.25$, that is close to the transition from type-I to type-II regimes and where we start obtaining vortices in our setups (Figure \ref{fig:VortexLowerLimitPlot}). This suggests that, just like the transition from type-I to type-II migration speed is modulated by the gap depth (eq. \eqref{eq:type-ItoIIMigTransition}), so should the eccentricity damping. We now look for a more quantitative expression of this fact, and since the data is much cleaner in panel (b), we use the observed gap-depth rather than the predicted one. We will deal later on with the question of how to predict a gap depth in an $N$-body setting without resorting to computationally expensive hydrodynamical simulations.

For low eccentricities, $0<e/h\lesssim0.4$, one can very well fit $1/\tau_e$ vs the observed gap depth with a straight line over the full gap depth range considered here ($\Sigma_{\min}/\Sigma_0 \simeq 0.3$ to 1). In the limit of no gap ($\Sigma_{\min}/\Sigma_0\simeq 1$), which should correspond to the setup of \cite{2008A&A...482..677C}, the observed eccentricity damping efficiency is slightly higher than the prediction $1/\tau_{e, \mathrm{CN2008}}$ (or in other words, $\tau_e$ is slightly smaller than $\tau_{e, \mathrm{CN2008}}$). We note that we are using a 2D setup with very high resolution, while \cite{2008A&A...482..677C} used a 3D setup with lower resolution. Moreover, the hydrodynamical codes used are different, and \cite{2008A&A...482..677C} fitted $e$-damping timescales to evolving planets, while we keep the planet on a fixed orbit and extract $\tau_e$ from the forces felt by the planet from the disc. We also run cases with the same setup as \cite{2008A&A...482..677C} and found that we still obtain a slightly more efficient e-damping. The difference, however is smaller that 20\%, which is the accuracy of \cite{2011MNRAS.410..293P} torque formula anyway.
For higher eccentricities up to $e/h\simeq 1$ the points again follow a straight line for gap depths $\simeq 0.3$ up to $\simeq 0.8$, after which $e$-damping becomes super-linear and significantly more efficient than \cite{2008A&A...482..677C}'s prediction. We checked this result with \citealt{2000A&AS..141..165M}'s original \texttt{fargo} code, yielding very similar results to ours for the orbital damping timescales, with a difference of only 1\%. 
One explanation for this effect is that for shallower gaps, and thus thinner gaps too, and sufficiently high $e$, the planet's excursions around $r_\pl=a$ due to the eccentricity of the orbit start having a significant effect. Since the gap around a planet starts to be carved at $r_\pl\pm 2/3 H$ where the Lindblad torques accumulate, we see that the planet's orbits starts interacting more with the edge of the gap at  $e/h\simeq 1$. Similar effects were observed in \citealt{2010A&A...523A..30B} and \citealt{2014MNRAS.437...96F}.

\begin{figure}[t!]
\centering
\includegraphics[width=0.45\textwidth]{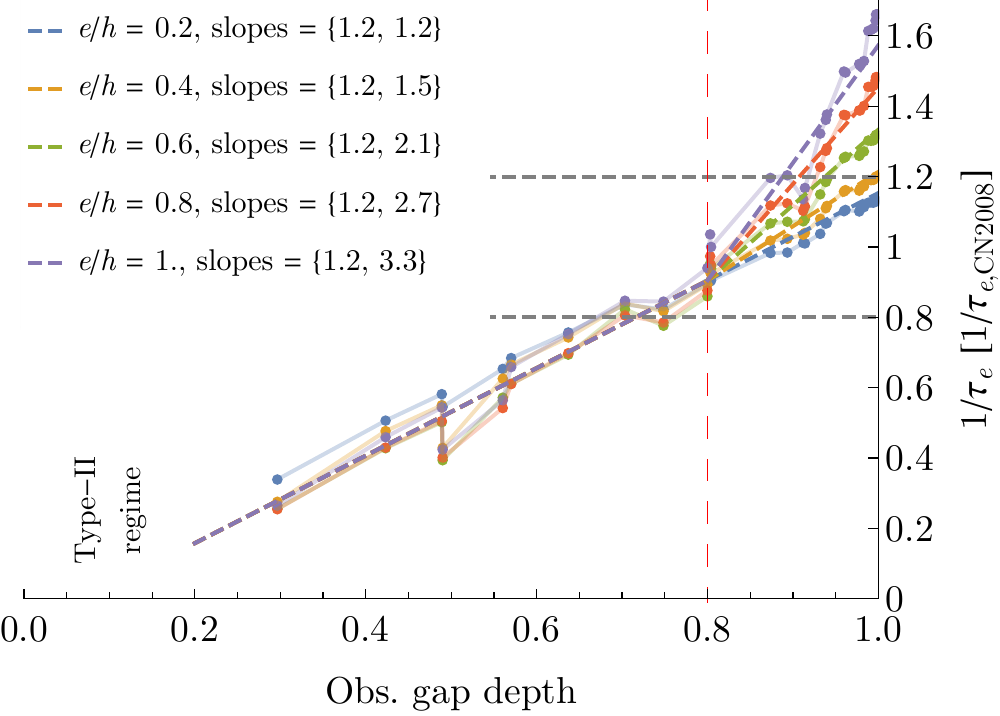}
\caption{Double linear fit of the $e$-damping efficiency to the data from panel (b) of Figure \ref{fig:edampVsGapDepth}, according to equation \eqref{eq:FitSegmented}. The fit to the data is split into a linear fit over gap depths $\in [0.3,0.8]$, and a separate linear fit over gap depths $\in [0.8,1]$ which continuously joins with the previous fit. The slopes for both fits are given in the legend in the top-left corner.}
\label{fig:edampVsObsGapDepthFitSegmented2}%
\end{figure}

We thus consider a fit to the data separating the horizontal axis into gap depths deeper and shallower than 80\%, and we perform a double fit over the two portions which joins continuously at gap depths of 80\% (Figure \ref{fig:edampVsObsGapDepthFitSegmented2}).
We find that the simple piecewise linear fit gives a very good approximation to the data:
\begin{equation}\label{eq:FitSegmented}
\begin{split}
    \frac{1}{\tau_e} &= \frac{1}{\tau_{e, \mathrm{CN2008}}}\times \begin{cases} 
    c_1 - m_1 \left(1 -\frac{\Sigma_{\min}}{\Sigma_0}\right)  & \text{if $\frac{\Sigma_{\min}}{\Sigma_0} < 0.8$}, \\
    c_2 - m_2 \left(1 -\frac{\Sigma_{\min}}{\Sigma_0}\right) & \text{if $\frac{\Sigma_{\min}}{\Sigma_0} > 0.8$}
    \end{cases}\\
    &\text{$m_1 = 1.263$, $c_1 = 1.169$, $m_2 = \tilde{m}$, $c_2 = 0.916 + 0.2 \tilde{m}$}
\end{split}
\end{equation}
where $\tilde{m} = \max\left\{m_1, m_1 + 3(e/h - 0.3)\right\}$. In particular, the multiplicative factor would in general depend on $e/h$, but its expression is rather simple. We show in Figure \ref{fig:edampVsObsGapDepthFitSegmented2} a comparison between the fit and the data. The typical error given by the fit is of the order of 4\%, and is overall less than 20\%, which is the accuracy of torque formulas from the literature.
When $\Sigma_{\min}/\Sigma_0=1$ (no measured gap) we recover \cite{2008A&A...482..677C} within errors equivalent to the error to the torque typically used in the literature \citep{2011MNRAS.410..293P}, while for deeper and deeper gaps formula \eqref{eq:FitSegmented} yields a simple model for the reduction of $e$-damping efficiency due to partial gap opening.

\section{Discussion}\label{sec:Discussion}

\subsection{Practical estimation of the gap profile}\label{subsec:PEGP}
The results of Section \ref{subsec:PartialGapResult} on the shape and depth of the gaps carved by planets indicate that a more careful analysis is needed when going to lower disc viscosities.  
In addition, for our own purposes of fitting $e$-damping timescales for use in $N$-body codes, the two panels from Figure \ref{fig:edampVsGapDepth} show that a better fit for the gap depths at lower viscosities is needed in order to make any practical use our equation \eqref{eq:FitSegmented}.
We recall that, down to $\alphaturb$-viscosities of about $10^{-3}$, \citealt{2018ApJ...861..140K}'s prediction gives relatively good agreements also with our numerical experiments (with a perfect fit at $\alphaturb=10^{-3}$, see Fig.\ \ref{fig:KanagawaVsObs_GapDepth}); moreover, formula \eqref{eq:KanagawaGapDepth} has the unquestionable advantage of being very easy to handle in practice for $N$-body codes and population synthesis studies.
Thus, we take a practical approach, and we build upon \citealt{2018ApJ...861..140K}'s ideas to obtain a similar fit to our data.
In particular, we rewrite \citealt{2018ApJ...861..140K}'s formula \eqref{eq:KanagawaGapDepth} as 
\begin{equation}\label{eq:KFromGapDepth}
    K = 25\left(\frac{1}{\Sigma_{\min}/\Sigma_0}-1\right),
\end{equation}
and we fit a function of the form $\tilde K(q,\alphaturb,h) = C q^{p_q} \alphaturb^{p_{\alphaturb}} h^{p_h}$ using our observed values for the right-hand-side of \eqref{eq:KFromGapDepth}. We used \texttt{Mathematica}'s \texttt{NonlinearModelFit} function and obtained the following fit:
\begin{equation}\label{eq:tildeK}
    \tilde K = 3.93 q^{2.3} h^{-6.14} \alphaturb^{-0.66}.
\end{equation}
\begin{figure}[t!]
\centering
\includegraphics[width= 0.45\textwidth ]{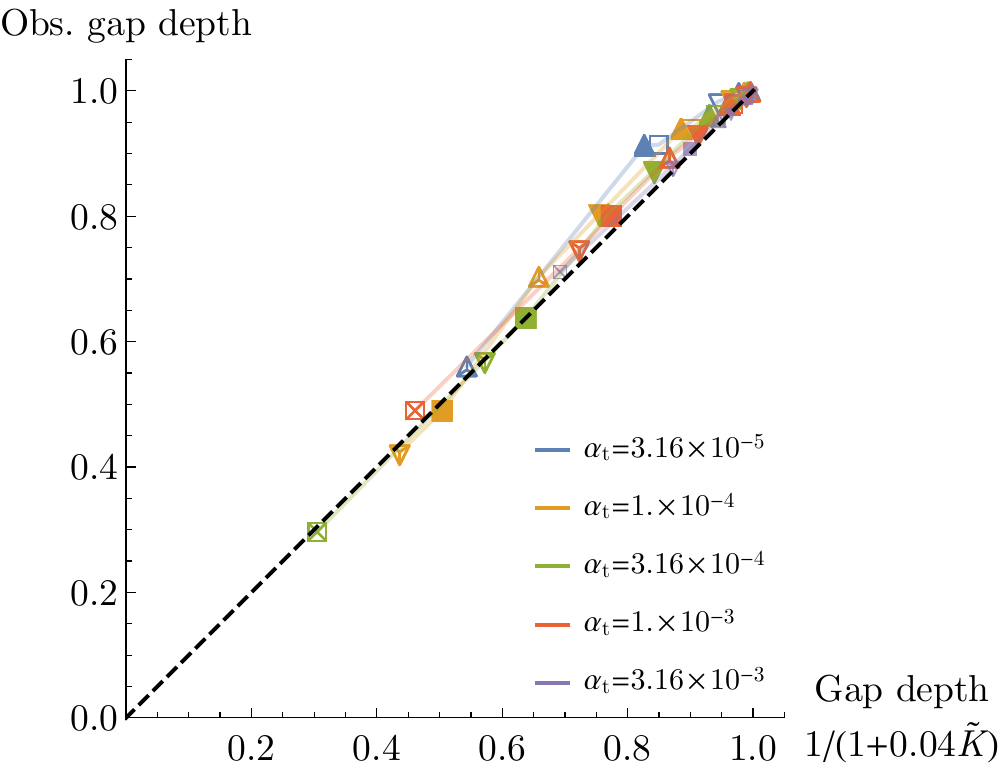}
\caption{Similar to Fig.\ \ref{fig:KanagawaVsObs_GapDepth}, but using \eqref{eq:TildeKanagawaGapDepth} as an approximation for the gap depth, showing better agreement with the data. As in Fig \ref{fig:KanagawaVsObs_GapDepth}, different colours represent different levels of viscosity as shown in the legend, and different symbols represent different aspect ratios and planetary masses: squares, downward-pointing triangles and upward-pointing triangles for $h=0.04$, $0.05$ and $0.06$ respectively; empty, filled and crossed symbols for $\mpl/M_*=10^{-5}$, $3\times 10^{-5}$ and $6\times 10^{-5}$ respectively.}
\label{fig:TildeKanagawaVsObs_GapDepth}%
\end{figure}
We show in Figure \ref{fig:TildeKanagawaVsObs_GapDepth} an analogue of Figure \ref{fig:KanagawaVsObs_GapDepth} but using \eqref{eq:tildeK}'s $\tilde K$ instead of \eqref{eq:KanagawaK}'s $K$, showing much better agreement.
This suggests a modification to the gap depth formula, namely
\begin{equation}\label{eq:TildeKanagawaGapDepth}
    \Sigma_{\min}/\Sigma_0\simeq \frac{1}{1+0.04 \tilde{K}}
\end{equation}

We should stress, however, that \eqref{eq:tildeK} is only a fit to the data, and does not emerge from a carefull study of the gas dynamics at low- to moderate-viscosities due to the presence of a planet. However, its sole purpose is to be combined with equation \eqref{eq:FitSegmented} in order to give a practical recipe for $e$-damping without resorting to hydro simulations (see next Section).
Moreover, we note that 2D and 3D simulations can be different regarding gap opening (see appendix of \citealt{2018A&A...612A..30B}), which has not been considered here.
A more detailed analysis on the matter will be the subject of future studies, as well as a study in the full 3D case.
We note that \cite{2014ApJ...782...88F} and \cite{2020A&A...643A.133B} already showed that the gap opening at low viscosities does not follow the predictions from higher viscosity simulations $\alphaturb \gtrsim 10^{-3}$. Also, one should always keep in mind that these studies have been carried out using different hydrodynamical codes.\\

\subsection{Implementation of eccentricity damping to $N$-body codes for partial gap-opening planets}
\begin{figure}[t!]
\centering
\includegraphics[width=0.45\textwidth]{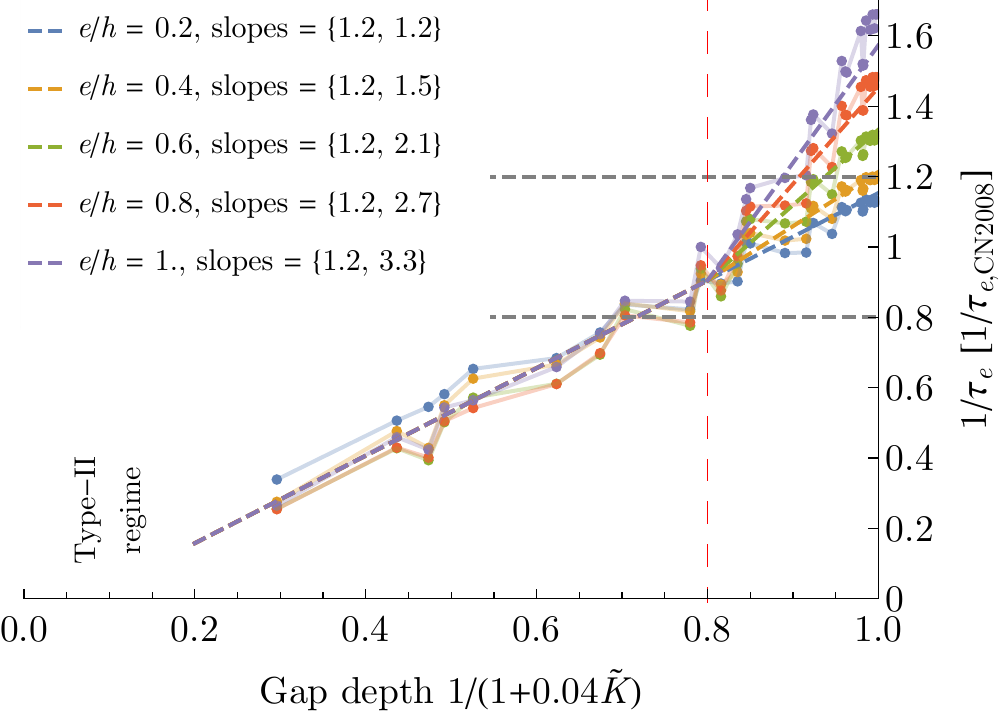}
\caption{Similar to Fig.\ \ref{fig:edampVsObsGapDepthFitSegmented2}, but using \eqref{eq:TildeKanagawaGapDepth} as an approximation for the gap depth on the horizontal axis. The fit to the $e$-damping efficiency obtained from hydro-dynamical simulations remains good, within the typical variations of analytical formulas for type-I migration.}
\label{fig:edampVsTildaKGapDepthFitSegmented}%
\end{figure}

Using then the approximate gap depth \eqref{eq:TildeKanagawaGapDepth} we obtain, together with \eqref{eq:FitSegmented}, a practical recipe to implement $e$-damping in low viscosity discs for partial-gap opening planets. The resulting comparison to the data, using \eqref{eq:TildeKanagawaGapDepth} instead of the observed gap depth like in Figure \ref{fig:edampVsObsGapDepthFitSegmented2}, is shown in Figure \ref{fig:edampVsTildaKGapDepthFitSegmented}. The typical error of this fit is still of the order of 4\%, and less than the $\simeq20\%$ of the torque formalisms. An example of comparison between the output of our hydro simulations and the fit is already shown in Figure \ref{fig:ComparisonPlot}.

Concerning the left-hand-side of Figure \ref{fig:edampVsTildaKGapDepthFitSegmented}, we recall that according to \citealt{2018ApJ...861..140K} the transition between type-I and type-II migration regimes should occur when the gap depth is about 50\% (see their Figure 4); instead, \citealt{2006Icar..181..587C} consider (arbitrarily) this transition should happen when the bottom density is $\lesssim 10\%$ the unperturbed surface density. Such deep gaps can be achieved for massive planets, but it is unclear whether a low-mass planet of a few (or a few tens) of Earth masses in a low viscosity disc would be able to carve such gaps without the emergence of a vortex. In any case, we saw in Section \ref{subsec:PGO} that a gap of order a few $10^{-1}$ is the threshold down to which one can consider the type-I regime valid. This means that we are able to probe reasonably deep gaps, even in the lowest viscosity cases considered here, down to the type-I to type-II transition, so that the results presented in this paper can be safely used in $N$-body implementations where a real disc is not evolved.
Intermediate gap depths seem to be the easiest to grasp: The $e$-damping efficiency is simply linear with the gap depth, with very little dependence on the eccentricity (see also Appendix \ref{apx:EEDRPG}). The $e$-independent fit reported in the first line of equation \eqref{eq:FitSegmented} captures the main effect of reduced $e$-damping efficiency for intermediate gaps with respect to the commonly used formula \eqref{eq:taue_CN2008} from \citealt{2008A&A...482..677C}; it is thus very similar to the expression for the change in migration timescale for partial-gap opening planets, equation \eqref{eq:type-ItoIIMigTransition}, which is the approach that has been typically used in the literature.
Concerning instead the right-hand side of Figure \ref{fig:edampVsTildaKGapDepthFitSegmented},  The deviation from the $e$-independent linear fit of at high eccentricities, as mentioned in Section \ref{subsec:eDampEffResult}, seems to be caused by the interaction with the edge of the gap. The latter is known to show differences in 2D vs.\ 3D hydro simulations. Moreover, there are known 2D and 3D differences for the change of eccentricity \citep{2010A&A...523A..30B}. Thus, some differences might be expected when going into the 3D case. 
This will be the subject of future work.

\subsection{Consequences for mean motion resonant locking via convergent migration}
There is strong theoretical and observational evidence that the dynamical histories of Super-Earth/Mini-Neptune systems have been shaped by a phase of type-I-migration driven capture in mean motion resonance \citep{2017MNRAS.470.1750I, 2021A&A...650A.152I}.
\cite{2019MNRAS.489L..17M}'s hydrodynamical simulations suggest instead that, in the limit of an inviscid disc, resonant capture might be avoided by planets that would otherwise build a resonant chain in a viscous disc. Their setup includes 5 planets of increasing mass with increasing separation from the central star, inside a 2D disc with a prescribed surface density and temperature profile, in the case of a viscous ($\alphaturb$ of order $10^{-3}$) and inviscid disc. While in the viscous case the stable systems were always found to be locked in mean motion resonances, the inviscid case did not form complete chains despite the final system being closely packed.
\cite{2019MNRAS.489L..17M} do not give a definitive explanation for the difference in outcome for viscous and inviscid discs. They mention that significant perturbations to the disc structure can occur in the inviscid case (such as vortices), which may be the reason why some planets escape mean motion resonances and undergo chaotic close encounters. 
It is however not easy to draw definitive conclusions on exactly what this outcome is due to, and whether it is representative for the bulk of the Super-Earth/Mini-Neptune population. Indeed, it is hard to determine a precise dynamical history from the evolution of a 5-planet system; also a setup with increasing planetary mass is in effect built to favour instabilities, since resonances tend to be less stable when the outer planet is more massive than the inner one \citep{2008A&A...478..929M, 2015ApJ...810..119D}. 
A completely inviscid disc might also not be a fully realistic environment, as even in the MRI-dead zone a residual turbulent viscosities can arise by purely hydrodynamical instabilities, with $\alphaturb$ of order $10^{-4}$ (e.g.\ \citealt{2021ApJ...915..130P, 2020ApJ...897..155F}), and observational constraints determine $\alphaturb$ in discs to range between $10^{-4}$ to a few $10^{-2}$ \citep{2017ApJ...837..163R, 2018ApJ...869L..46D, 2017ApJ...843..150F, 2018ApJ...856..117F}. Finally, we do observe a few systems in resonance, and their observed architecture has been used to determine {\it a-posteriori} the level of viscosity: in the case of Kepler-223, this indirect approach has yielded an estimate on $\alphaturb$ of the order of $10^{-3}$ \citep{2021A&A...656A.115H}.

In any case, in light of \cite{2019MNRAS.489L..17M}'s results, one should expect a transition from efficient to inefficient resonant chain assembly as planet-disc interactions become more and more significant even in the type-I regime, due to lower viscosity discs. 
Our results from Figure \ref{fig:edampVsGapDepth} might give an initial hint as to why this can be the case, even when vortices are not formed by these planets. Indeed, we found that the eccentricity damping provided by the disc can be significantly less effective for deeper and deeper gaps. During resonant capture, the orbital eccentricities are pumped by the planet-planet interaction until an equilibrium configuration is obtained between resonant pumping and disc-driven $e$-damping \citep{2018CeMDA.130...54P}. Less efficient $e$-damping thus means higher orbital eccentricities, which lead to a more excited state more prone to instabilities. Finally, the emergence of vortices may indeed play a role as \cite{2019MNRAS.489L..17M} suggest. 
Clearly more work is needed in order to obtain a more detailed picture of type-I-like interactions for low-viscosity discs in the context of shaping the exoplanet population and for applications to population synthesis models.

We conclude remarking the work of \citealt{2021A&A...648A..69A}, who performed 2D locally isothermal simulations of resonant trapping for low ($\alphaturb= 5\times10^{-5}$) viscosity discs in a variety of configurations, implementing two different types of disc inner edge to stop the inner planet's migration. Since they also considered increasing planetary masses for further-out planets, some simulations failed to build chains, remarking that the more massive the outer planet is, the earlier the system becomes unstable, which is in line with the aforementioned $N$-body and theoretical investigations on the subject. Still, their simulations with 2 -- 3 planets were able in many cases to form resonant chains that remained stable also after the removal of the gas.
Indeed, resonant systems of 2 to 3 planets are known to be relatively stable for low-mass planets and low eccentricities \citep{2018CeMDA.130...54P, 2020MNRAS.494.4950P}. For higher number of planets or planetary cores, a less efficient $e$-damping in low viscosity (but not inviscid) discs allows for chains to form during the disc phase which, because of the increased eccentricities with respect to classical type-I migration $e$-damping, become more susceptible to instabilities after the dissipation of the disc. This would have consequences in population synthesis works, such as the so-called ``breaking the chains'' scenario \citep{2017MNRAS.470.1750I, 2021A&A...650A.152I}, allowing instabilities to develop also for planets of lower masses than the ones needed in these works to efficiently break the resonant chains.

\section{Conclusions}\label{sec:Conclusion}
In this paper, we have revisited the problem of orbital eccentricity damping timescales for planets that are in the type-I migration regime but which open partial to significant gaps, for example due to the low viscosity in their surrounding disc. These planets, typically in the Super-Earth/Mini-Neptune mass range, are therefore not expected to undergo type-II migration, but their interaction with thin, low-viscosity discs can be viewed as pertaining to a transition regime between type-I and type-II migration. In general, the shape and depth of the gap carved by the planet and the resulting change in migration timescales (i.e.\ the torque felt by the planet) in such intermediate cases (up to and including classical type-II interactions) have been the subject of many theoretical and numerical works (e.g.\ \citealt{2006Icar..181..587C, 2011MNRAS.410..293P, 2017MNRAS.471.4917J, 2018ApJ...861..140K}); these theories have been implemented in multiple population synthesis works (e.g.\ \citealt{2017MNRAS.470.1750I, 2018MNRAS.474..886N, 2019A&A...623A..88B, 2020ApJ...892..124O, 2021A&A...650A.152I, 2021A&A...656A..69E}) to reproduce the observed characteristics of exoplanetary systems. In particular, the transition in migration regimes from type-I and type-II has been described as a simple reduction to the type-I torque proportional to the gap-depth carved by the planet \citep{2018ApJ...861..140K}.
However, a similar study for the change in eccentricity damping has not been performed. Instead, the aforementioned population synthesis works have used a fixed $e$-damping efficiency following the work of \citealt{2008A&A...482..677C}, which fitted the evolution of Super-Earths embedded in a disc of relatively high viscosity ($\alphaturb=5\times10^{-3}$) to obtain orbital damping timescales.
These high $\alphaturb$ values might however not be the norm in planet forming regions, meaning that even such low-mass planets as the ones considered in \citealt{2008A&A...482..677C} would open partial gaps at lower viscosities ($\alphaturb\lesssim10^{-4}$). Besides, it is known that the orbital eccentricities can even be pumped by planet-disc interactions for planets that have carved a deep-enough gap \citep{2001A&A...366..263P, 2006A&A...447..369K, 2013A&A...555A.124B}. We thus expect there to be a transition between ``pure'' type-I eccentricity damping \emph{à la} \citealt{2008A&A...482..677C} and less efficient $e$-damping up to even $e$-pumping for planets that carve intermediate to deep gaps.\\

We thus explored this transition making use of high resolution, 2D, locally isothermal hydro-dynamical simulations of Super-Earth-type planets of varying masses ($\mpl/M_* = 1$ to $6 \times 10^{-5}$) and varying orbital eccentricities ($e/h = 0$ to 1, where $h=H/r$ is the disc's aspect ratio), embedded in discs of varying viscosities ($\alphaturb = 10^{-3}$ down to $3.16\times10^{-5}$) and aspect ratios ($h = 0.04$ to $0.06$). The planet is kept on a fixed orbit and the system is evolved for thousands of the planet's orbital period until a steady-state is achieved, all the while recording the planet-disc interaction forces that drive migration and $e$-damping. The $e=0$ case is used in order to measure gap depths $\Sigma_{t_{\max}}/\Sigma_0$, where $\Sigma_{t_{\max}}$ is the radial  surface density profile at the end of the simulation and $\Sigma_0$ is the initial, unperturbed surface density (a power law). We compared our observed gap depths with the already existing results from \citealt{2018ApJ...861..140K}, showing good agreement at the higher viscosities considered by \citealt{2018ApJ...861..140K} (down to $\alphaturb\simeq 10^{-3}$) but less so for the lower viscosities; in particular, their formula gave an over-estimate to the gap depths observed in our own simulations. 

We thus found in Section \ref{subsec:PEGP} a fit to the gap depth similar to that of \citealt{2018ApJ...861..140K}'s, which gives better agreement with our low-viscosity simulation, as a function of the discs aspect ratio and viscosity and the planet's mass.
Then, we recorded the observed $e$-damping efficiency for the various planetary eccentricities, as a function of the observed gap depth. For deep gaps, a clear linear trend is observed, with $e$-damping efficiency scaling with the gap depth and being as low as 1/4 the expected value from \citealt{2008A&A...482..677C}'s prescription at the transition between type-I and type-II migration. The result is thus similar to the case of diminished migration efficiency for partial-gap opening planets proposed in \citealt{2018ApJ...861..140K} and used extensively in the aforementioned population synthesis works. For shallow gaps, the same simple linear trend remains for low eccentricities ($e/h\lesssim0.2$), while we observe an increase in eccentricity damping efficiency for higher eccentricities. This can be understood as due to the interaction of the eccentric planet with strongly non-axisymmetric structures in the disc, in particular with the gap carved by the planet itself. 

A simple $e$-dependent fit is found for all cases in equation \eqref{eq:FitSegmented}, which yields typical errors smaller than or comparable to the accuracy of widely-used analytical type-I migration prescriptions \citep{2011MNRAS.410..293P}. Together with the gap-depth fit \eqref{eq:TildeKanagawaGapDepth}, this allows us to provide a simple prescription for eccentricity-damping in type-I regime for partial-gap opening planets that is consistent with high-resolution 2D hydro-dynamical simulations.
3D simulations as well as more sophisticated thermodynamics to relax the locally isothermal assumption will be the subject of future work, in order to obtain a full picture for eccentricity and inclination damping efficiency for partial gap opening planets.

\begin{acknowledgements}
G.\ P.\ and B.\ B.\ thank the European Research Council (ERC Starting Grant 757448-PAMDORA) for their financial support.
We acknowledge HPC resources  from "Mesocentre SIGAMM" hosted by Observatoire de la C\^ote d'Azur. L.\ E.\ wish to thank Alain Miniussi for maintenance and re-factorisation of the code FARGOCA.
G.\ P.\ wishes to thank Frédéric Masset for helpful discussions. 
The authors wish to thank the anonymous referee for helpful comments and suggestions.
\end{acknowledgements}

%
%

\begin{appendix}
\section{Disc surface density and torque response}\label{apx:TypicalDiscEvo}
\begin{figure*}[ht!]
   \resizebox{\hsize}{!}
        {
            \subfigure[$t$ = 25 orbits]{
            \includegraphics[width=.3\textwidth]{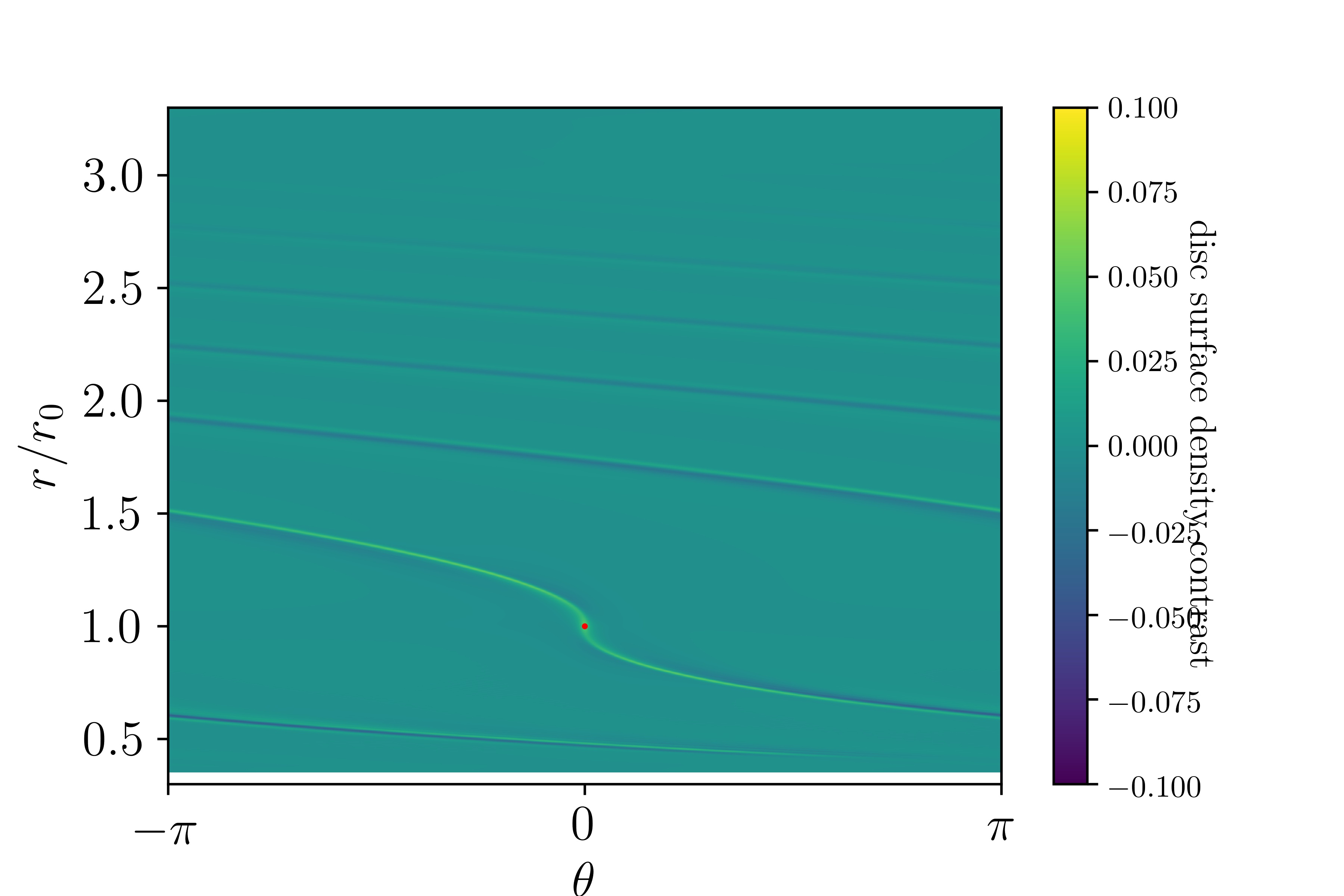}
            }%
            \subfigure[$t$ = 1000 orbits]{
            \includegraphics[width=.3\textwidth]{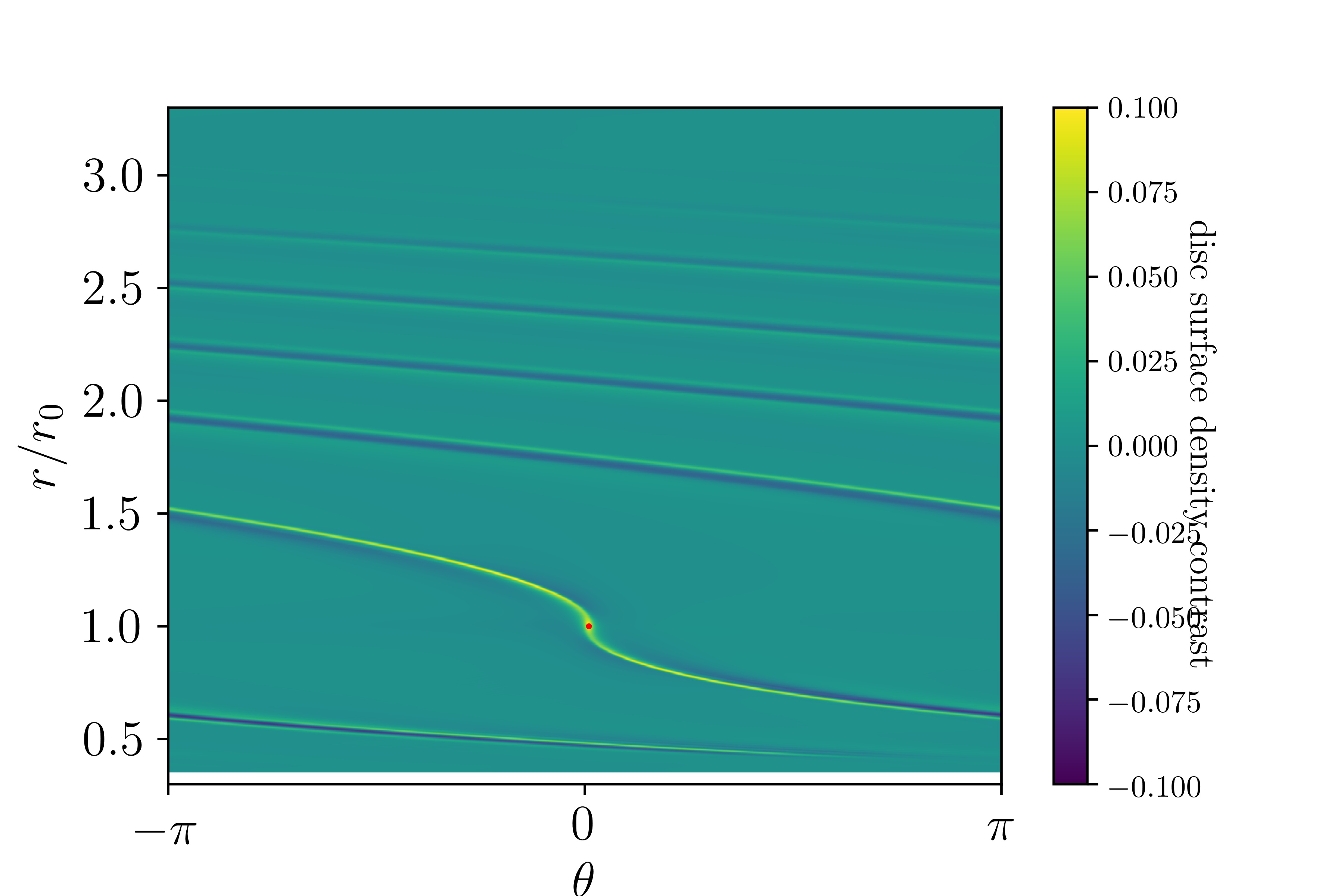}
            }%
            \subfigure[$t$ = 2000 orbits]{
            \includegraphics[width=.3\textwidth]{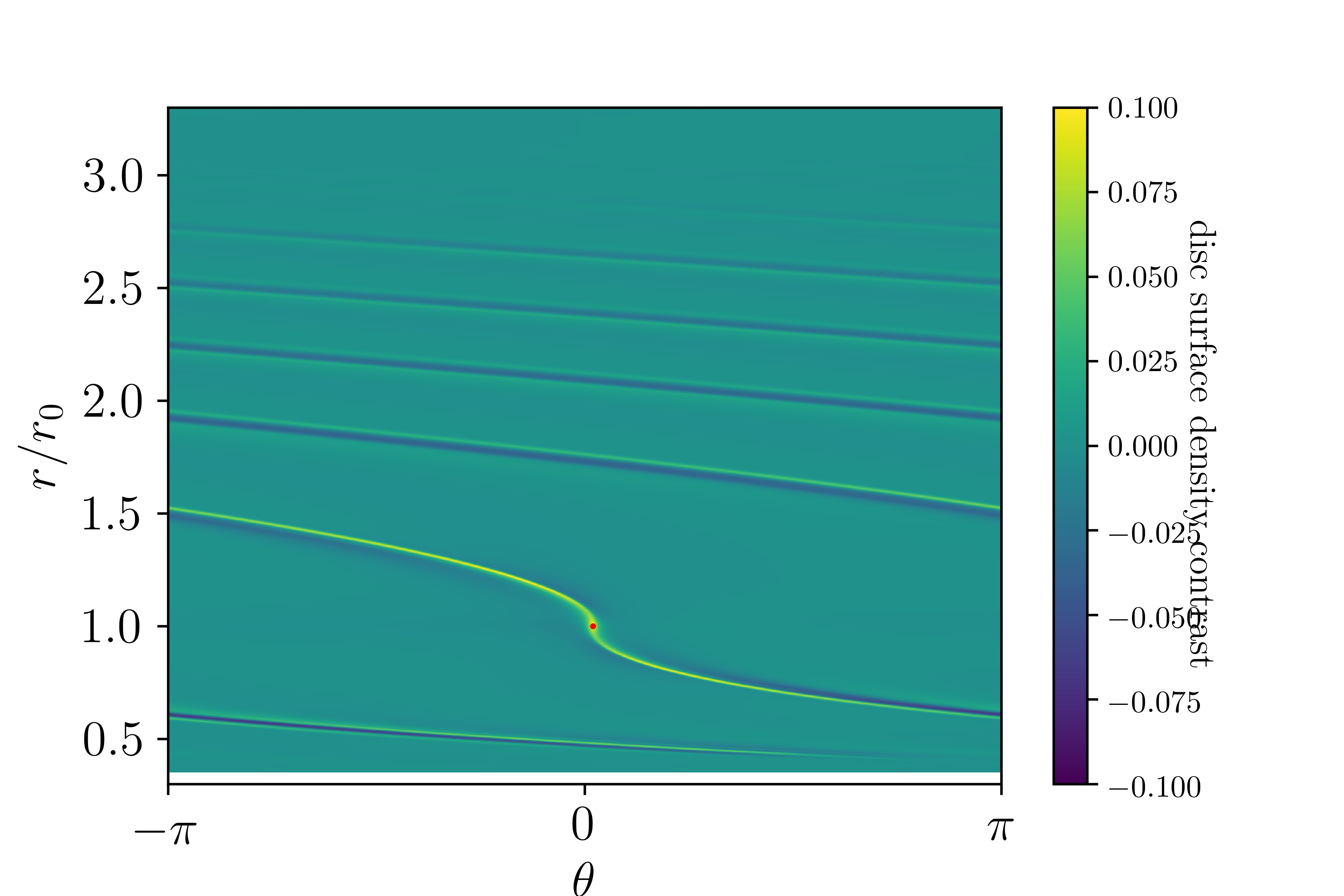}
            }%
        }
      \caption{Surface density evolution in a setup with $\alphaturb = 10^{-3}$ and $h=0.05$, over 3000 orbits of a $\mpl/M_* = 1\times 10^{-5}$ planet on a (fixed) circular orbit at 1 AU. We show the surface density contrast 
      $(\Sigma_t(r,\theta)-\left<\Sigma_t(r)\right>_\theta)/\left<\Sigma_t(r)\right>_\theta$ on the $(\theta,r)$ plane, where $\left<\Sigma_t(r)\right>_\theta$ is the azimuthally averaged surface density at the distance $r$.
      }
\label{fig:NoVortex}
\end{figure*}
\begin{figure}[t!]
\centering
\includegraphics[width= 0.45\textwidth ]{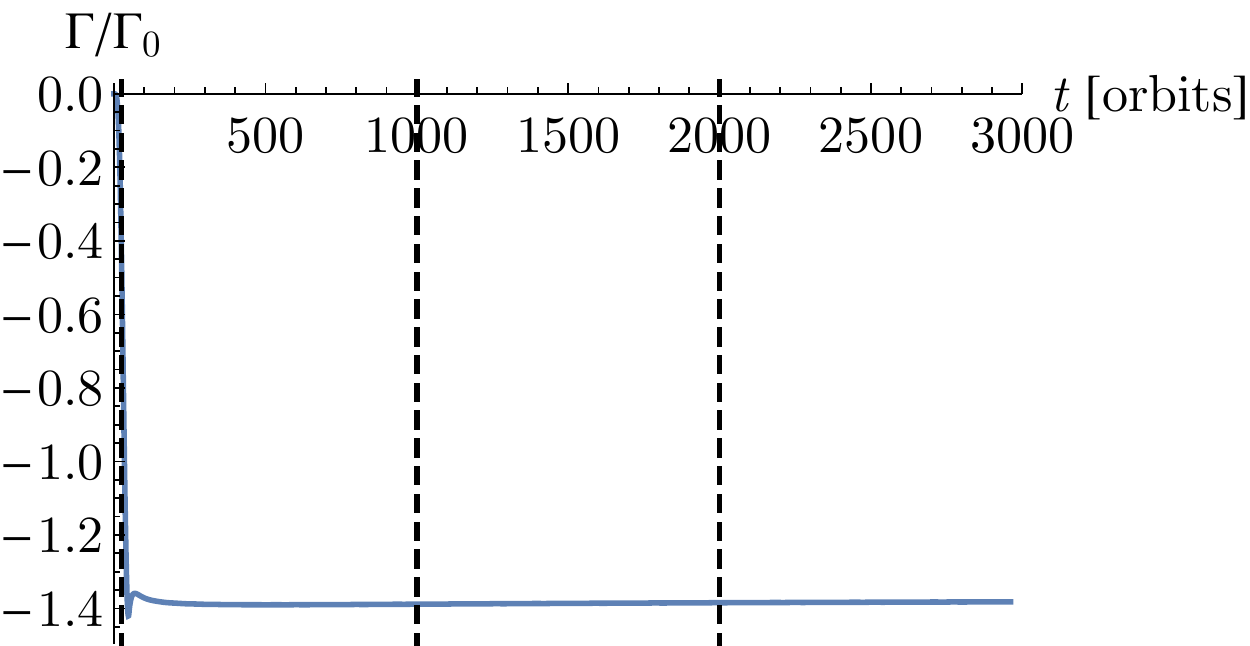}
\caption{Evolution of the torque felt by the planet in the setup presented in Figure \ref{fig:NoVortex}, as a function of time. Vertical dashed line represent timestamps for which the surface density is shown in Figure \ref{fig:NoVortex}. The torque stabilises quickly in this case, as the system reaches a long-lived steady state.}
\label{fig:NoVortex_Torque}%
\end{figure}
\begin{figure*}[ht!]
   \resizebox{\hsize}{!}
        {
            \subfigure[]{
            \includegraphics[width=0.45\textwidth]{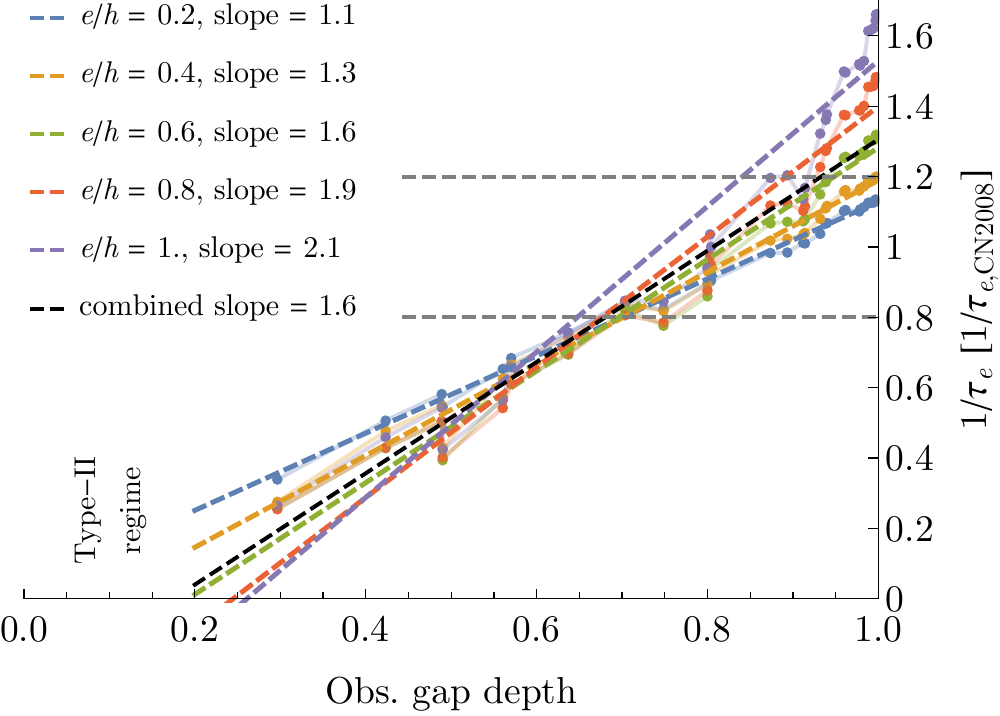}
            }%
            \subfigure[]{
            \includegraphics[width=0.45\textwidth]{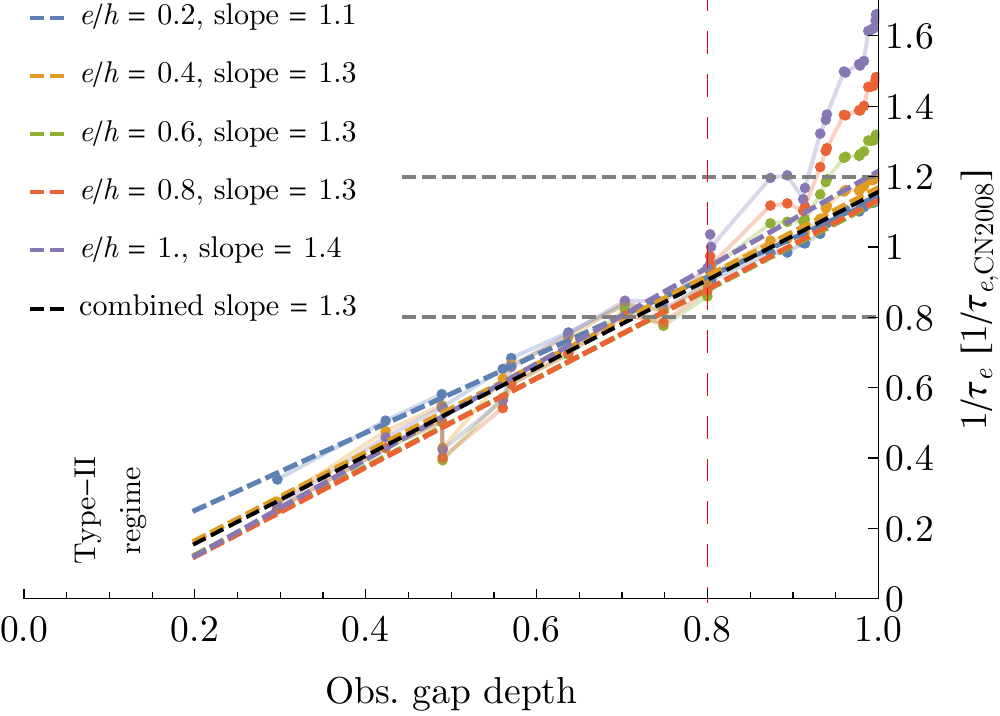}
            }%
        }
      \caption{Linear fit of the $e$-damping efficiency to the data from panel (b) of Figure \ref{fig:edampVsGapDepth}, i.e.\ using the gap depths observed from the simulations on the horizontal axis. In panel (a) we fit the full data, both for the seperate values of $e/h$ (coloured dashed lines) and for the full data set combining all eccentricities together (black dashed line). In panel (b) we fit the data in the same way, but considering gap depths of up to 80\% (indicated by the dashed red vertical line). In both cases we report the observed slopes of the lines in the top-left legend.
      }
\label{fig:edampVsGapDepthFit}
\end{figure*}
We present here snapshots from a typical evolution of one of our discs, with $\alphaturb = 10^{-3}$ and $h=0.05$, interacting with a $\mpl/M_* = 1\times 10^{-5}$ planet. Figure \ref{fig:NoVortex} depicts the surface density contrast, showing the emergence of the wake typical of type-I response of the disc due to the presence of the planet; Figure \ref{fig:NoVortex_Torque} shows the stabilisation of the torque felt by the planet from the disc.

\section{Estimate of eccentricity-damping and pumping resonances for partial gaps}\label{apx:EEDRPG}

For an eccentric planet, the Fourier decomposition of the gravitational planet-disc potential includes terms that are proportional to the eccentricity and which therefore drive its evolution. Typically, for a small planet that does not significantly perturb the disc profile, the eccentricity is both damped and pumped by these individual so-called first-order Lindblad resonances (although their combination usually results in damping of the eccentricity, see below). 
The main resonances that damp $e$ are the co-orbital first-order Lindblad Resonances with $|l-m|=1$ 
(see \citealt{1980ApJ...241..425G, 1988Icar...73..330W, 2008EAS....29..165M}; see also the Appendix of \citealt{2015ApJ...812...94D} for a summary). These are located at the location of the planet, i.e.\ $\Omega = \Omega_\mathrm{pl}$, and their strength decreases with $m$.
The main resonances that drive the eccentricity are the ``External'' first-order Lindblad Resonances. These are located interior (for $l=m+1$, $\Omega=\left(\frac{m+1}{m-1}\right)\Omega_\mathrm{pl}$) or exterior (for $l=m-1$, $\Omega=\left(\frac{m-1}{m+1}\right)\Omega_\mathrm{pl}$) to the planet's orbit, get closer and closer to the planet's orbit for increasing $m$, and their sizes also initially increases with $m$, but ultimately become smaller for higher $m$'s.

For small bodies (which do not open a gap), \citealt{1993ApJ...419..166A} found that the $e$-damping due to the co-orbital Lindblad resonances overcomes any excitation due to the external resonances by roughly a factor of 3; instead, for massive enough planets, their gap is significantly deep at the co-rotation radius that the $e$-damping co-orbital resonance's effect is quenched. Sufficiently massive planets can even have their eccentricities even excited \citep{2001A&A...366..263P, 2006A&A...447..369K, 2013A&A...555A.124B}, due to the combined effect of the remaining resonances (e.g.\ \citealt{2015ApJ...812...94D}).
Thus, as an intermediate-mass planet carves a partial gap, we expect to transition smoothly from one regime ($e$-damping) to the other ($e$-pumping).

One might imagine that, as a gap is cleared at the location of the planet, the efficiency of $e$-damping resonances decreases in proportion to the decrease of gas surface density, thus explaining the decrease in $e$-damping efficiency.
However, works on the shape and width of gaps (\citealt{2006Icar..181..587C, 2018ApJ...861..140K}, as well as our own simulations) show that as soon as a gap opens, it already has a finite width, which remains relatively constant as the gap's depth increases. 
Thus, as material is removed at the location of the $e$-damping resonances (i.e.\ the co-rotation radius) so is the case for some $e$-pumping resonances (the ``external'' Lindblad Resonances closest to the planet), so the picture is not immediately clear. However, we computed the effect of each specific first-order Lindblad Resonances using disc profiles from our hydro-simulation: we find that this effect is almost completely balanced by the over-density at the inner and outer edges of the gap, while at the same time the locations closer to the planets are associated to higher indices $m$ which already did not contribute much to the final $\dot{e}$. Thus, the resulting $e$-pumping efficiency remains relatively constant and only decreases by less than 10\%.
Since the main term that governs $\dot{e}/e$ is given by the $e$-damping terms corresponding to the the co-orbital first-order Lindblad Resonances with $l = m + 1, m-1$, and these are proportional to the gap depth in surface density, this explains why, down to a decrease of $e$-damping efficiency by roughly a factor of a few due to the carving of a partial gap at the location of the planet, we expect this decrease to be linear with ${\Sigma_{\min}}/{\Sigma_0}$.

\section{Analysis of indirect forces}
In our setup, the planets' orbits are fixed and the forces from the disc to the planet are recorded. However, this situation is not purely consistent, since slight disc asymmetries can cause a non-vanishing acceleration onto the star, meaning that the reference frame with the star at the origin is non-inertial. These indirect forces might thus have an effect on the osculating elements, and therefore also on the eccentricity damping rates. 
In order to assess the effect of indirect forces, we track the acceleration onto the star due to its interaction with the disc. Since again the star is fixed onto the origin in the code's reference frame, this results in an equal but opposite acceleration onto the planet. We see however that this acceleration is, on average, of the order of about two orders of magnitude smaller than the acceleration due to direct effects between the disc and the planet.
We thus conclude that the asymmetries of the disc generated by the planet's effect onto the disc itself are not large enough to cause large indirect forces contribution to the planet's evolution. The picture would change in the case of more massive planets and more pronounced disc asymmetries.

\section{Fits to $e$-damping efficiency}
We report in this appendix additional analysis on the data described in Section \ref{subsec:eDampEffResult}. Figure \ref{fig:edampVsGapDepthFit}(a) shows $e$-dependent (coloured dashed lines) and global (black dashed line) linear fits to the observed $e$-damping efficiency along the full gap depth axis. While for small values of $e/h$ this yields a good fit to the data, we see that this is not so for higher eccentricities, $e/h\sim 1$. The global fit (black dashed line) fails to accurately reproduce the data within acceptable errors for type-I migration formulas. For this reason, in Section \ref{subsec:eDampEffResult} we resort to a double-linear fit that is eccentricity dependent. The left-part of the fit is shown in Figure \ref{fig:edampVsGapDepthFit}(b) for clarity. Here, the data used to make the fit has been truncated for observed gap depths of at least 80\%. In this case, the combined fit alone (black dashed line) is able to fully recover the data within reasonable variations of less than 10\%. The fit in Section \ref{subsec:eDampEffResult} represents an extension of this approach, where an $e$-dependent fit to the right-hand side of the plot (gap depths shallower than 80\%) is continuously connected to the simple $e$-independent fit for deeper gaps.

\end{appendix}

\end{document}